\documentclass[a4paper,11pt]{article}
\pdfoutput=1

\usepackage{amsmath}
\usepackage{amssymb}
\usepackage[small]{caption}
\usepackage{cite}
\usepackage{enumerate}
\usepackage[margin=1in]{geometry}
\usepackage[multiple]{footmisc}
\usepackage{graphicx}
\usepackage{physics}
\usepackage{slashed}
\usepackage{textcomp}
\usepackage[nottoc]{tocbibind}
\usepackage{xcolor}

\definecolor{mediumblue}{rgb}{0,0,0.8}
\usepackage{hyperref}
\hypersetup{
  linktocpage=true,
  colorlinks=true,
  citecolor=mediumblue,
  filecolor=mediumblue,
  linkcolor=mediumblue,
  urlcolor=mediumblue,
}
\newcommand{\mailref}[1]{\href{mailto:#1}{#1}}

\providecommand{\keywords}[1]
{
  \small
  \textbf{Keywords---} #1
}

\usepackage[T1]{fontenc}
\usepackage{listings}
\definecolor{dkgreen}{rgb}{0,0.6,0}
\definecolor{gray}{rgb}{0.5,0.5,0.5}
\definecolor{mauve}{rgb}{0.58,0,0.82}
\definecolor{bggray}{RGB}{247,247,247}
\usepackage{varwidth}

\graphicspath{{./}{./figures/}}
\numberwithin{equation}{section}
\allowdisplaybreaks%

\DeclareMathOperator*{\argmin}{argmin}
\newcommand{\CC}{C\nolinebreak\hspace{-.05em}\raisebox{.4ex}{\tiny\bf +}\nolinebreak\hspace{-.10em}\raisebox{.4ex}{\tiny\bf +}}

\def\thefootnote{\fnsymbol{footnote}}

\begin{document}

\begin{titlepage}
  \begin{flushright}
    CTPU-PTC-20-18
  \end{flushright}

  \bigskip

  \begin{center}
    \bf \LARGE
    YAM2: Yet another library for the \boldmath{$M_2$} variables\\[2mm]
    using sequential quadratic programming
  \end{center}

  \medskip

  \begin{center}
    \bf \large Chan~Beom~Park\footnote{E-mail: \texttt{\mailref{cbpark@ibs.re.kr}}}
  \end{center}

  \begin{center}
    \em Center for Theoretical Physics of the Universe,
    Institute for Basic Science (IBS),\\
    55 Expo-ro, Yuseong-gu, Daejeon 34126, Korea\\[0.2cm]
  \end{center}

  \medskip

  \begin{abstract}
    The $M_2$ variables are devised to extend $M_{T2}$ by promoting
    transverse masses to Lorentz-invariant ones and making explicit
    use of on-shell mass relations.
    Unlike simple kinematic variables such as the invariant mass of
    visible particles, where the variable definitions directly provide
    how to calculate them, the calculation of the $M_2$ variables is
    undertaken by employing numerical algorithms.
    Essentially, the calculation of $M_2$ corresponds to solving a
    constrained minimization problem in mathematical optimization, and
    various numerical methods exist for the task.
    We find that the sequential quadratic programming method performs
    very well for the calculation of $M_2$, and its numerical
    performance is even better than the method implemented in the
    existing software package for $M_2$.
    As a consequence of our study, we have developed and released yet
    another software library, \texttt{YAM2}, for calculating the $M_2$
    variables using several numerical algorithms.
  \end{abstract}

  \medskip

  \noindent




\end{titlepage}

\renewcommand{\thefootnote}{\arabic{footnote}}
\setcounter{footnote}{0}

\setcounter{tocdepth}{2}
\noindent\rule{\textwidth}{0.3pt}\vspace{-0.4cm}\tableofcontents
\noindent\rule{\textwidth}{0.3pt}

\section{Introduction}

\noindent
The existence of invisible particles such as neutrino or dark matter
candidate in the final state of physics processes poses a great
challenge on physic analyses at hadron colliders, where the event
information along the beam direction is limited, in particular.
Detectors at collider experiments project the full phase space onto
the space of visible particle momenta, and the projection is a
non-invertible mapping.
We can infer the existence of invisible particles only by the record
of missing transverse momentum, or equivalently, the momentum
imbalance of visible final-state particles.
In response to the challenge, physicists have developed many useful
methods and algorithms for extracting the information of the physics
processes involved with the invisible particle as much as possible.
As one of such endeavors, a set of kinematic variables, termed $M_2$,
has been proposed for measuring the mass spectrum in the presence of
invisible particles produced in a pair and providing a good
approximation to the four-momenta of the invisible
particles~\cite{Cho:2014naa}.

The $M_2$ variables are an extension of $M_{T2}$~\cite{Lester:1999tx,
  Barr:2003rg} to the Lorentz invariant systems without projecting on
the transverse plane.
In the definition of $M_2$, the longitudinal momenta, as well as the
transverse momenta of invisible particles, become the parameters of
minimization under certain kinematic constraints.
The value and the solution of $M_2$ are obtained from the optimality
and feasibility conditions of the nonlinearly constrained minimization
problem in mathematical optimization.
The formulation and description of the $M_2$
variables will be given in Sec.~\ref{sec:M2}.

Unlike simple kinematic variables such as the invariant mass of
visible particles, where the variable definitions directly provide how
to calculate them, the calculation of the $M_2$ variables is
undertaken by employing numerical algorithms.
It is due to the lack of analytic expressions for $M_2$, except for
some special kinematic configurations.
Currently, the only publicly available software package for
calculating the $M_2$ variables is
\texttt{OPTIMASS}~\cite{Cho:2015laa}.

As mentioned in the above, finding the $M_2$ value is to perform
constrained minimization in essence, and there are a lot of numerical
methods viable for the task.
We will present a brief overview of some of such methods, the
augmented Lagrangian method and the sequential quadratic programming,
in Sec.~\ref{sec:algo}.
We have attempted to implement the numerical methods for comparing
their performance of the $M_2$ calculation.
As an outcome of our study, we release yet another software library
\texttt{YAM2} for calculating $M_2$.
The implementation of numerical algorithms in \texttt{YAM2} and the
benchmark study of comparing the algorithms are presented in
Sec.~\ref{sec:performance}.
We find that the sequential quadratic programming method
performs better than the other methods, including those implemented in
\texttt{OPTIMASS}.
The instructions for installing and using \texttt{YAM2} are described
in Sec.~\ref{sec:install}, and then the last section is dedicated to
summary and outlook.

\section{\label{sec:M2}The \boldmath{$M_2$} variables}

\noindent
In the physics process under consideration, if there exist particles
moving beyond the outermost detector component without leaving a
trace, it is recorded as the signal with missing energy.
Reconstruction of missing energy events is one of the major obstacles
to physics analyses at hadron colliders.
The reconstruction is an attempt of inverse projection onto the full
phase space, which is bound to be incomplete.
Nevertheless, in many situations, it is possible to measure the mass
spectrum or to reconstruct the missing energy events up to some
ambiguity.
One of the multitudes of methods aiming to resolve the missing energy
problem is the method of the $M_2$ variables~\cite{Cho:2014naa,
  Cho:2015laa, Kim:2017awi}.

Before entering into the description of the $M_2$ variables, it would
be better to start with looking into $M_{T2}$, which is closely
related to the definition of the $M_2$ variables and designed to
tackle the same problem.
The invention of the $M_{T2}$ variable was devised to
find a lower bound on the masses of superparticles decaying into
the lightest neutral supersymmetric particle.
Such examples include the pair
productions of sleptons, $\widetilde \ell^+ + \widetilde \ell^- \to
\ell^+ \widetilde\chi_1^0 + \ell^- \widetilde\chi_1^0$~\cite{Lester:1999tx},
charginos, $\widetilde\chi_1^+ + \widetilde\chi_1^- \to \pi^+
\widetilde\chi_1^0 + \pi^- \widetilde\chi_1^0$~\cite{Barr:2002ex,
  Barr:2003rg},
and gluinos, $\widetilde g + \widetilde g \to q \bar q \widetilde\chi_1^0
+ q \bar q \widetilde\chi_1^0$~\cite{Cho:2007qv, Barr:2007hy,
  Cho:2007dh, Nojiri:2008vq}.
All of them fall into the symmetric decay topology of
\begin{equation}
  Y + \bar Y \longrightarrow v_1 (p_1) \chi (k_1) + v_2 (p_2)\bar\chi
  (k_2),
  \label{eq:pair_proc}
\end{equation}
where $v_i$ are the sets of visible standard model particles, and
$\chi$ is the invisible particle.
The method of the $M_{T2}$ variable is indeed applicable to not only
supersymmetric cases but any physics process that can be represented
by the decay topology~(\ref{eq:pair_proc}).
It is defined as
\begin{align}
  M_{T2} \equiv
  &~ \min_{\vb*{k}_{1T}, \, \vb*{k}_{2T} \in \mathbb{R}^2}
  \Big[ \max \Big\{ M_{T} \left( p_{1T},
      \, k_{1T};\, M_\chi \right), \,
    M_{T} \left( p_{2T}, \, k_{2T};\, M_\chi \right) \Big\} \Big]
    \nonumber\\
  &~ \text{subject to}\,\,\, \vb*{k}_{1T} + \vb*{k}_{2T} =
  \slashed{\vb*{P}}_T ,
  \label{eq:MT2}
\end{align}
where $\slashed{\vb*{P}}_T$ is the missing transverse momentum determined
by the negative sum over all the visible particles momenta in the plane
transverse to the beam axis,
\begin{equation}
  \slashed{\vb*{P}}_T = - \sum_i \vb*{p}_{iT} .
\end{equation}
If there are additional particles not involved with the hard process,
such as initial state radiations, they are added to the sum as well.
The invisible particle mass $M_\chi$, which is unknown, is an input
for the transverse masses,
\begin{equation}
  M_T^2 (p_{iT}, \, k_{iT};\, M_\chi) =
  \pi_T \left( p_i + k_i \right)^2 .
  \label{Eq:MT}
\end{equation}
Here $\pi_T$ is the projection operator from ($1 + 3$)-dimensional
space onto ($1+2$)-dimensional one:
\begin{align}
  \begin{split}
    & \pi_T:\, p_i = \left(E_i, \, \vb*{p}_i \right)
    \rightarrow p_{iT} = \left( E_{iT} ,\, \vb*{p}_{iT} \right), \\
    & \pi_T:\, k_i = \left(e_i, \, \vb*{k}_i \right)
    \rightarrow k_{iT} = \left( e_{iT} ,\, \vb*{k}_{iT} \right) ,
  \end{split}
      \label{eq:proj_transverse}
\end{align}
with $E_{iT} = (m_i^2 + \norm{\vb*{p}_{iT}}^2)^{1/2}$ and $e_{iT} =
(M_\chi^2 + \norm{\vb*{k}_{iT}}^2)^{1/2}$.
Note that the transverse masses in (\ref{Eq:MT}) are convex functions
over the invisible transverse momenta $\vb*{k}_{iT}$ because their
Hessian matrices are positive semi-definite~\cite{Lim:2016ymd}:
\begin{equation}
  \det \left[ \mathbf{H} M_T^2 (p_{iT}, \, k_{iT};\, M_\chi) \right] =
  \frac{4 E_{iT}^2 M_\chi^2}{e_{iT}^4} \geq 0 .
\end{equation}
One can further find that taking the maximum between the two
transverse masses does not violate the convexity property.
Therefore, any local minimum found by the $M_{T2}$ calculation is
automatically a global minimum.

The $M_{T2}$ distribution has an endpoint at the parent particle mass
$M_Y$ for the true value of $M_\chi$, that is,
\begin{equation}
  M_{T2} (M_\chi = M_\chi^\text{true}) \leq M_Y.
\end{equation}
Thus, it enables us to extract the mass spectrum information by
identifying the position of the endpoint of the $M_{T2}$ distribution.

It does not stop there: the solution of the minimization
in~(\ref{eq:MT2}) provides an approximation to the transverse momenta
of invisible particles,
\begin{equation}
  \left\{ \widetilde{\vb*{k}}_{iT} \right\}
  = \argmin_{\vb*{k}_{1T} + \vb*{k}_{2T} = \slashed{\vb*{P}}_T}
  \Big[ \max \Big\{ M_{T} \left( p_{1T},
      \, k_{1T};\, M_\chi \right), \,
      M_{T} \left( p_{2T}, \, k_{2T};\, M_\chi \right) \Big\} \Big] .
      \label{eq:MT2_sol}
\end{equation}
Then, for the approximation to the transverse momenta
$\widetilde{\vb*{k}}_{iT}$, one can obtain the associated longitudinal
momenta $\widetilde{k}_{iL}$ of the invisible particles by using the
on-shell mass relations of the parent particles:
\begin{equation}
  ( p_1 +  \widetilde k_1 )^2 = M_Y^2 , \quad
  ( p_2 +  \widetilde k_2 )^2 = M_Y^2 .
\end{equation}
It is called the $M_{T2}$-assisted on-shell (MAOS) method or the MAOS
approximation for the invisible momenta~\cite{Cho:2008tj}.
The right-hand sides of the equations in the above need not be the
parent particle mass.
One can substitute the $M_{T2}$ value or the transverse mass $M_T$ in place
of $M_Y$~\cite{Choi:2009hn, Cho:2009wh, Choi:2010dw, Park:2011uz}.
In this way, the MAOS method serves an approximate reconstruction of
the center-of-mass frame event by event. It can be used to
measure the particle properties, such as spins and helicities, besides
the mass spectrum information.

Except for some special cases~\cite{Lester:2007fq, Barr:2007hy,
  Cho:2007dh, Burns:2008va, Cho:2009wh, Konar:2009qr, Agashe:2010tu,
  Lester:2011nj, Lally:2012uj, Mahbubani:2012kx}, the analytic
expression for the $M_{T2}$ variable and the
solution~(\ref{eq:MT2_sol}) in general cases are unknown.
Instead, one makes use of numerical optimization algorithms to
calculate them.\footnote{
  In this article, the term ``optimization'' is interchangeable
  with ``minimization.''
}
Note that $M_{T2}$ can be written as
\begin{equation}
  M_{T2} =
  \min_{\vb*{k}_{1T} \in \mathbb{R}^2} \Big[ \max \Big\{ M_{T} \left( p_{1T},
      \, k_{1T};\, M_\chi\right), \,
      M_{T} \left( p_{2T}, \, (e_{2T}, \, \slashed{\vb*{P}}_T -
        \vb*{k}_{1T});\, M_\chi \right) \Big\} \Big] ,
      \label{eq:MT2_2}
\end{equation}
where $\vb*{k}_{2T}$ has been eliminated by the constraint on the
missing transverse momentum, and $e_{2T} =$ $(M_\chi^2 +
  \norm{\slashed{\vb*{P}}_T - \vb*{k}_{1T}}^2)^{1/2}$.
Therefore, finding the $M_{T2}$ value for a given event corresponds to
performing {\em unconstrained} minimization on a function of two variables,
$\vb*{k}_{1T}$ $=(k_{1x}$, $k_{1y})$.
In the past, a combination of Migrad and Simplex methods
included in the \texttt{Minuit2} library~\cite{James:1975dr} of
\texttt{ROOT} was used for the minimization~\cite{Lester:MT2}.
The Migrad algorithm is a variable-metric method that depends on the
first derivative of the objective function to be
minimized~\cite{10.1093/comjnl/6.2.163, Davidon:1966mm}, while the
Simplex algorithm, also called the Nelder-Mead
method~\cite{Nelder:1965zz}, does not use the derivative information.
For a brief review of the methods, see Appendix~A of
Ref.~\cite{Cho:2015laa}.
Afterward, it was realized that $M_{T2}$ could be understood as the
boundary of the mass region ($M_\chi$, $M_Y$) consistent with the minimal
kinematic constraints, {\em i.e.}, the on-shell mass relations,
\begin{equation}
  (p_1 + k_1)^2 = (p_2 + k_2)^2 = M_Y^2,
  \label{eq:on_shell_MY}
\end{equation}
and the missing transverse momentum constraint~\cite{Cheng:2008hk}.
The kinematically allowed region for $\vb*{k}_{1T}$ forms an ellipse,
and $M_{T2}$ can be calculated by investigating the scaling behavior
of two ellipses for the two decay chains in (\ref{eq:pair_proc}).
In the implementation, it is used the bisection method in conjunction with
the Sturm sequence for the quartic polynomial, converted from two
quadratic equations, to test if two ellipses intersect.
See Subsec.~2.3 and Appendix A of Ref.~\cite{Cheng:2008hk} for the detail.
The coded implementation is faster and more accurate than the
implementation based on the Migrad and Simplex algorithms, so it has
served as the {\em de facto} standard calculator for $M_{T2}$ for both
theoretical and experimental analyses.
There is an alternative calculator, which is also based on the
bisection method, but with higher precision and less numerical
instabilities than the previous one~\cite{Lester:2014yga}.

Now we turn our attention to the $M_2$ variables.
Astute readers may notice that the subscript ``$T$'' has been dropped as
compared with $M_{T2}$. It is indeed a ($1+3$)-dimensional analogue of
$M_{T2}$ without projecting the particle momenta onto the transverse
plane as in~(\ref{eq:proj_transverse}).
For the decay topology~(\ref{eq:pair_proc}), it is defined by
\begin{align}
  M_{2} \equiv
  &~ \min_{\vb*{k}_{1}, \, \vb*{k}_{2} \in \mathbb{R}^3}
  \Big[ \max \Big\{ M \left( p_{1},
      \, k_{1};\, M_\chi \right), \,
    M \left( p_{2}, \, k_{2};\, M_\chi \right) \Big\} \Big]
    \nonumber\\
  &~ \text{subject to}\,\,\, \vb*{k}_{1T} + \vb*{k}_{2T} =
    \slashed{\vb*{P}}_T ,
    \label{eq:M2}
\end{align}
where ${M(p_i, \, k_i)}^2 = {(p_i + k_i)}^2$.
It is six-dimensional constrained minimization over $\vb*{k}_1$ and
$\vb*{k}_2$.
Each invariant mass is a convex function over the corresponding
invisible momenta since its Hessian matrix is positive semi-definite,
\begin{equation}
  \det \left[ \mathbf{H} M^2 (p_{i}, \, k_{i};\, M_\chi) \right] =
  \frac{8 E_{i}^3 M_\chi^2}{e_{i}^5} \geq 0 .
\end{equation}
Therefore, the objective function of $M_2$ is also convex, and it is
sufficient to search for a local solution to obtain the $M_{2}$ value.

As we have seen in Eq.~(\ref{eq:MT2_2}), we can
eliminate the components $\vb*{k}_{2T}$ by using the missing
transverse momentum constraint.
Thus, finding the $M_2$ value corresponds to performing
four-dimensional {\em unconstrained} minimization in essence.
The unknown variables are $k_{1x}$, $k_{1y}$, $k_{1z}$, and $k_{2z}$.
The invisible particle mass $M_\chi$ is again an input.
Due to the minimization, the $M_2$ distribution is also bounded from
above by the parent particle mass $M_Y$.
Despite the increased number of variables for minimization, it is
found that $M_2$ defined in Eq.~(\ref{eq:M2}) is equivalent to $M_{T2}$:
they have the same value event by event~\cite{Cho:2014naa,
  Barr:2011xt}.
It is because the minimization over $k_{1z}$ and $k_{2z}$ results in the
vanishing of rapidity differences, $\Delta \eta_i = \eta_{p_i} - \eta_{k_i}
= 0$, where the invariant masses $M (p_{i}$, $k_{i})$ become
identical to the transverse masses $M_T (p_{iT}$, $k_{iT})$.

There are additional considerations worth examining further.
If the physics process has the decay
topology~(\ref{eq:pair_proc}), the longitudinal momenta $k_{iz}$
satisfy the on-shell mass relations for the parent particles, as
given in~(\ref{eq:on_shell_MY}).
Thus, we can impose the on-shell mass relation as a constraint in
addition to the missing transverse momentum constraint,
\begin{align}
  M_{2C} \equiv
  &~ \min_{\vb*{k}_{1}, \, \vb*{k}_{2} \in \mathbb{R}^3}
  \Big[ \max \Big\{ M \left( p_{1},
      \, k_{1};\, M_\chi \right), \,
    M \left( p_{2}, \, k_{2};\, M_\chi \right) \Big\} \Big]
    \nonumber\\
  &~ \text{subject to}\,\, \left\{\,
    \begin{aligned}
      \vb*{k}_{1T} + \vb*{k}_{2T}
      &= \slashed{\vb*{P}}_T , \\
      (p_1 + k_1)^2
      &= (p_2 + k_2)^2 .
    \end{aligned} \right .
        \label{eq:M2C}
\end{align}
Here the subscript ``$C$'' denotes the additional constraint on the
invariant masses.
We do not use $M_Y$ since it is unknown and is to be determined.
It is similar to the constrained mass variable in
Refs.~\cite{Ross:2007rm, Barr:2008ba}, where a constraint on the mass
difference $M_Y - M_\chi$ has imposed further.
After eliminating $\vb*{k}_{2T}$ by using the missing transverse
momentum as before, it has the form of {\em constrained} minimization
over four variables.
However, it again turned out to be that $M_{2C}$ is equivalent to
$M_{T2}$~\cite{Cho:2014naa, Mahbubani:2012kx}.
Recall that $M_{T2}$ can be defined as the boundary of the consistent
mass region, subject to the kinematic constraints, including the
on-shell mass relations for the parent particles
in Eq.~(\ref{eq:on_shell_MY}).
In other words, $M_{T2}$ already utilizes the on-shell mass
information in an implicit way. Together with the fact that $M_{2} =
M_{T2}$, we also find that $M_{2C} = M_{2}$.

The $M_2$ variable becomes distinct from $M_{T2}$ when taking into
account on-shell intermediate particles in the decay chains:
\begin{align}
  A_1 + A_2
  &\longrightarrow a_1 \, B_1 + a_2 \, B_2 \nonumber\\
  &\longrightarrow a_1 (p_{a_1})\, b_1 (p_{b_1})\, C_1 (k_1)
    + a_2 (p_{a_2})\, b_2 (p_{b_2})\, C_2 (k_2).
    \label{eq:pair_cascade}
\end{align}
In the final state, $a_i$ and $b_i$ are visible particles, and $C_i$
are invisible particles responsible for the missing energy. $B_i$
are the intermediate states decaying to $b_i C_i$.
As in Ref.~\cite{Cho:2014naa}, we assume that the decay chains are
symmetric, {\em i.e.}, $M_{A_1} = M_{A_2}$ $= M_A$, $M_{B_1} = M_{B_2}$ $=
M_B$, and $M_{C_1} = M_{C_2}$ $= M_C$.
For the sake of notational simplicity, we express the visible momenta
by
\begin{equation}
  p_i \equiv p_{a_i} + p_{b_i}, \quad
  q_i \equiv p_{b_i} .
\end{equation}
Note that we have already used the relation that $k_1^2 = k_2^2 =
M_C^2$ since $M_C$ enters as an input to the invariant masses of
visible $+$ invisible particle systems in $M_2$. Then, the remaining
on-shell mass constraints yet to be used for the decay
topology~(\ref{eq:pair_cascade}) are
\begin{equation}
  (p_1 + k_1)^2
  = (p_2 + k_2)^2  , \quad
  (q_1 + k_1)^2
  = (q_2 + k_2)^2  .
\end{equation}
Depending on the on-shell mass constraint to use (or not to use)
for minimization, there are four types of the $M_2$ variables:
\begin{align}
  M_{2XX} \equiv
  &~ \min_{\vb*{k}_{1}, \, \vb*{k}_{2} \in \mathbb{R}^3}
  \Big[ \max \Big\{ M \left( p_{1},
      \, k_{1};\, M_C \right), \,
    M \left( p_{2}, \, k_{2};\, M_C \right) \Big\} \Big]
    \nonumber\\
  &~ \text{subject to}\,\,\,
    \vb*{k}_{1T} + \vb*{k}_{2T} = \slashed{\vb*{P}}_T,
    \label{eq:M2XX}\\
  M_{2CX} \equiv
  &~ \min_{\vb*{k}_{1}, \, \vb*{k}_{2} \in \mathbb{R}^3}
  \Big[ \max \Big\{ M \left( p_{1},
      \, k_{1};\, M_C \right), \,
    M \left( p_{2}, \, k_{2};\, M_C \right) \Big\} \Big]
    \nonumber\\
  &~ \text{subject to}\,\, \left\{\,
    \begin{aligned}
      \vb*{k}_{1T} + \vb*{k}_{2T}
      &= \slashed{\vb*{P}}_T , \\
      (p_1 + k_1)^2
      &= (p_2 + k_2)^2 ,
   \end{aligned} \right . \\
  M_{2XC} \equiv
  &~ \min_{\vb*{k}_{1}, \, \vb*{k}_{2} \in \mathbb{R}^3}
  \Big[ \max \Big\{ M \left( p_{1},
      \, k_{1};\, M_C \right), \,
    M \left( p_{2}, \, k_{2};\, M_C \right) \Big\} \Big]
    \nonumber\\
  &~ \text{subject to}\,\, \left\{\,
    \begin{aligned}
      \vb*{k}_{1T} + \vb*{k}_{2T}
      &= \slashed{\vb*{P}}_T , \\
      (q_1 + k_1)^2
      &= (q_2 + k_2)^2 ,
   \end{aligned} \right .\\
  M_{2CC} \equiv
  &~ \min_{\vb*{k}_{1}, \, \vb*{k}_{2} \in \mathbb{R}^3}
  \Big[ \max \Big\{ M \left( p_{1},
      \, k_{1};\, M_C \right), \,
    M \left( p_{2}, \, k_{2};\, M_C \right) \Big\} \Big]
    \nonumber\\
  &~ \text{subject to}\,\, \left\{\,
    \begin{aligned}
      \vb*{k}_{1T} + \vb*{k}_{2T}
      &= \slashed{\vb*{P}}_T , \\
      (p_1 + k_1)^2
      &= (p_2 + k_2)^2, \\
      (q_1 + k_1)^2
      &= (q_2 + k_2)^2 .
    \end{aligned} \right .
\end{align}
In each definition, the first subscript of $M_{2}$ refers to the
on-shell mass constraint on the parent particles $A_i$, and the second
does to that on the intermediate particles $B_i$.
The subscript ``$C$'' (``$X$'') means that the
corresponding constraint is (not) imposed.
One can also construct the $M_2$ variables for other subsystems of the
visible particles, {\em e.g.}, the subsystem of $\{ b_i \}$. Here we
confine ourselves to the $\{a_i b_i\}$ system.
One can see that $M_{2XX} = M_2$ in~(\ref{eq:M2}), and $M_{2CX} =
M_{2C}$ in~(\ref{eq:M2C}). Therefore, we have
\begin{equation}
  M_{2XX} = M_{2CX} = M_{T2} .
\end{equation}
On the other hand, it has been found that $M_{2XC}$ and $M_{2CC}$ are
different from $M_{T2}$, and further, they have the following
hierarchy~\cite{Cho:2014naa}:
\begin{equation}
  M_{2XX} = M_{2CX} \leq M_{2XC} \leq M_{2CC} \leq M_Y
  \label{eq:M2_hierarchy}
\end{equation}
for events with the decay topology~(\ref{eq:pair_cascade}).
The last inequality holds for $M_C = M_C^\text{true}$.
The consequence of the hierarchical structure is that the event
densities of the $M_{2XC}$ and $M_{2CC}$ distributions will be
populated more toward the parent particle mass than that of $M_{T2}$.

\begin{figure}[tb!]
  \centering
    \includegraphics[width=0.45\textwidth]{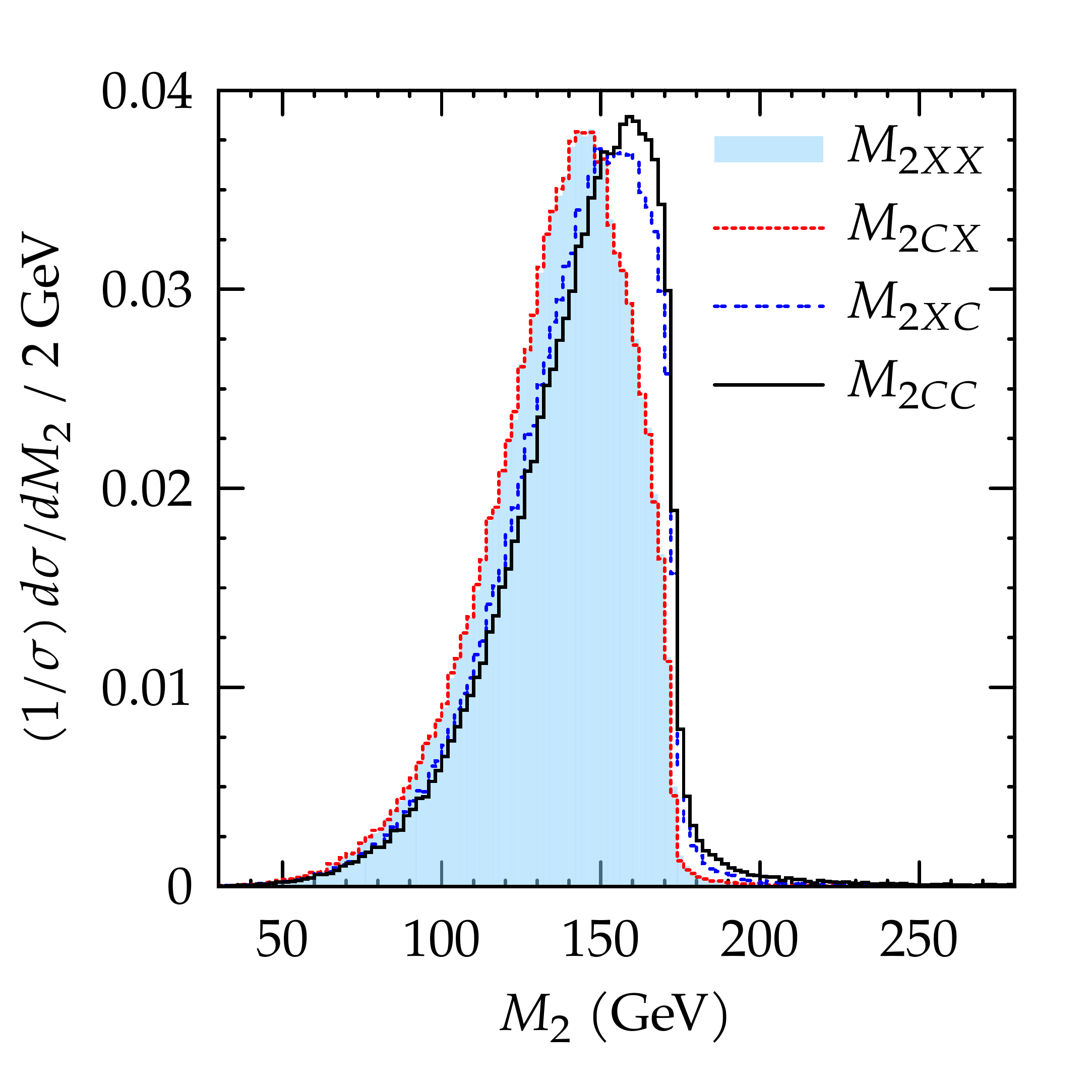}
    \includegraphics[width=0.45\textwidth]{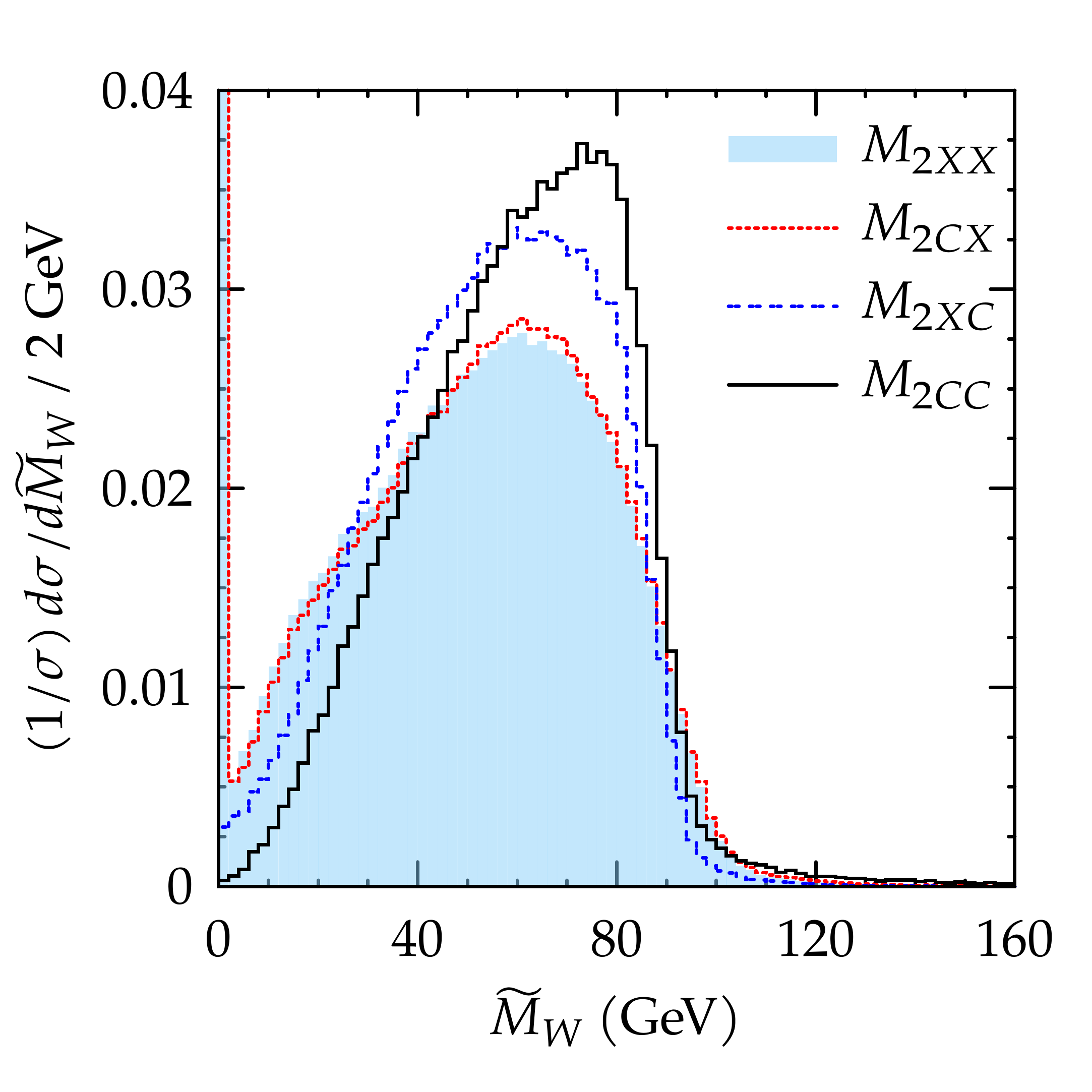}
  \caption{\label{fig:M2}
    The $M_2$ distributions (left) and the reconstructed $M_W$ using
    the $M_2$ solutions to the invisible neutrino momenta (right) for
    the di-leptonic top pair process at truth level. The description
    of the algorithm to obtain $M_2$ is given in
    Sec.~\ref{sec:performance}.}
\end{figure}

In the left panel of Fig.~\ref{fig:M2}, we show the $M_2$
distributions for the di-leptonic top pair process
\begin{equation}
  t + \bar t \to b W^+ + \bar b W^-
  \to b \ell^+ \nu + b \ell^- \bar \nu
  \label{eq:ttbar}
\end{equation}
at truth level with the LHC beam condition.\footnote{
  There is a combinatorial ambiguity on pairing the $b$ quarks and
  charged leptons. In this article, we do not concern the ambiguity,
  but use the correct pair. Interestingly, the $M_2$ variables can be
  used to resolve the combinatorial problem~\cite{Debnath:2017ktz}.
}
All distributions of the $M_2$ variables have endpoint shape near the
parent particle mass $m_t$, and the distributions of $M_{2XX}$ and
$M_{2CX}$ are identical to each other.
In particular, we can confirm the hierarchy
relation~(\ref{eq:M2_hierarchy}) by seeing the peak locations.

Another important outcome of the $M_2$ variables is that we can obtain
an approximation for the invisible particle momenta as the consequence
of the minimization:
\begin{align}
  \left\{ \widetilde{\vb*{k}}_i \right\}
  =&~ \argmin_{\vb*{k}_{1}, \, \vb*{k}_{2} \in \mathbb{R}^3}
  \Big[ \max \Big\{ M \left( p_{1},
      \, k_{1};\, M_C \right), \,
     M \left( p_{2}, \, k_{2};\, M_C \right) \Big\} \Big] \nonumber\\
  &~ \text{subject to constraints}.
\end{align}
It appears to be taking a similar approach to the MAOS method: {\em cf.}
Eq.~(\ref{eq:MT2_sol}). However, the methodology is substantially
different.
In the MAOS method, one obtains the transverse components of the
invisible momenta from the $M_{T2}$ solution and then employs the
on-shell mass relations to get the longitudinal components.
On the other hand, in the $M_2$ variables, the on-shell mass relations
act as the constraints in the minimization step, which eventually emits
the components of invisible momenta all together.
In result, the invisible momentum solutions of $M_{2XC}$ and
$M_{2CC}$ are uniquely determined, while each longitudinal momentum of
the MAOS method is determined up to two-fold ambiguity.
The comparison study of the $M_2$ and MAOS methods has been performed
in Ref.~\cite{Kim:2017awi}, where it is shown that the $M_2$ variables
provide the better approximation to the invisible momenta.
In the right panel of Fig.~\ref{fig:M2}, we show the reconstructed
$W$-boson mass using the $M_2$ solution to the neutrino momenta,
\begin{equation}
  \widetilde M_W^2 = (q_i + \widetilde{k}_i)^2.
\end{equation}
The peak of the $\widetilde M_W$ distribution for $M_{2CC}$ is located
near the $W$-boson mass, thus allowing additional mass measurement,
that is, measuring $m_W$ using $\widetilde M_W$ as well as $m_t$ using $M_{2}$.

\section{\label{sec:algo}Algorithms for constrained minimization}

\noindent
As in the case of $M_{T2}$, the calculation of the $M_2$ variables
resorts to numerical algorithms because analytic expression for
general cases is unknown.
The calculations for $M_2$ are essentially multi-dimensional
constrained minimization, except for $M_{2XX}$ in Eq.~(\ref{eq:M2XX}).
Currently, the only publicly available software package for
calculating $M_2$ is \texttt{OPTIMASS}~\cite{Cho:2015laa}.
The algorithm of choice in \texttt{OPTIMASS} is the augmented
Lagrangian (AUGLAG) method~\cite{Hestenes:1969, Powell:1969} with the
Migrad and Simplex algorithms from the \texttt{Minuit2} library of
\texttt{ROOT}.
In this section, we present a brief review of the formulation of
constrained minimization problems and numerical algorithms for
calculating the $M_2$ variables.

A general formulation for constrained minimization problems is
\begin{align}
  \min_{\vb*{x} \in \mathbb{R}^n} f(\vb*{x})
  \quad \text{subject to}\, \left\{
  \begin{aligned}
    c_i(\vb*{x}) &= 0 , \\
    d_j(\vb*{x}) &\geq 0 ,
  \end{aligned} \right . \label{eq:const_min}
\end{align}
where $f(\vb*{x})$ is an objective function, and $c_i (\vb*{x})$ and
$d_j (\vb*{x})$ are equality and inequality constraints, respectively.
The objective function and constraints are all smooth, real-valued
functions: $f$, $c_i$, $d_j$: $\mathbb{R}^n \to \mathbb{R}$.
In the case of $M_2$ variables, the objective function is
\begin{equation}
  f(\vb*{k}) =
  \max \Big\{ M \left( p_{1}, \, k_{1};\, M_C \right), \, M \left( p_{2},
    \, k_{2};\, M_C \right) \Big\} ,
  \label{eq:M2_obj}
\end{equation}
and $c_i$ are on-shell mass relations. The $M_2$ variables do not have
inequality constraints, $j \in \varnothing$.\footnote{
  If the decay width of unstable particles $A_i$ or $B_i$ are too
  large, we may have to include inequality constraints on the
  mass relations. In this article, we assume that the decay widths are
  negligible.
}
The unknown variables are $k_{1x}$, $k_{1y}$, $k_{1z}$, and $k_{2z}$,
after eliminating $k_{2x}$ and $k_{2y}$ by using the missing
transverse momentum condition.
If a certain point $\vb*{x}^\ast$ is in the feasible set for the
system of constraints,
\begin{equation}
  \vb*{x}^\ast \in \Omega
  = \big\{ \vb*{x} \in \mathbb{R}^n \,\vert\, c_i(\vb*{x}) = 0, \,
    d_j(\vb*{x}) \geq 0 \big\},
\end{equation}
and $f(\vb*{x}) \geq f(\vb*{x}^\ast)$ for a neighborhood of
$\vb*{x}^\ast$ on the feasible set $\Omega$, $\vb*{x}^\ast$ is termed
a local solution (or minimizer) of the problem.
For the $M_2$ variables, the $M_2$ value corresponds to
$f(\widetilde{\vb*{k}}_i)$, with $\widetilde{\vb*{k}}_i$ being the
$M_2$ solution to the invisible momenta event-by-event.
We refer to Ref.~\cite{Nocedal:2006} for the more complete
explanations of the constrained optimization problems and conditions
for the local solution.

As mentioned earlier, the software package \texttt{OPTIMASS} employs
the AUGLAG method, which is  well known and widely used algorithm for
constrained optimization problems. The AUGLAG method constructs a
Lagrangian function on top of the quadratic penalty function, in which
the penalty terms are the squares of constraints.
For the description of the AUGLAG method,
we begin by considering the penalty method for solving constrained
optimization problems.
Ignoring the inequality constraints, the quadratic penalty function
for the problem~(\ref{eq:const_min}) is
given by
\begin{equation}
  Q(\vb*{x};\, \mu) = f(\vb*{x}) + \frac{\mu}{2} \sum_i c_i^2(\vb*{x}) ,
\end{equation}
where $\mu > 0$ is the penalty parameter. The constraint violations,
or infeasibility, are penalized by increasing $\mu$.
We can minimize the penalty function $Q(\vb*{x};\, \mu)$ instead of
the objective function $f(\vb*{x})$.
By constructing the penalty function, the problem (\ref{eq:const_min})
has been transformed from {\em constrained} to {\em unconstrained}
minimization.
In the implementation of the algorithm, we increase the $\mu$ value,
and then seek the solution $\vb*{x}^\ast$ of $Q(\vb*{x};\, \mu)$ in
each iteration.
The iterations proceed until some convergence test has been satisfied.
However, it turns out that the solution $\vb*{x}^\ast$ does not
satisfy the feasibility conditions $c_i(\vb*{x}^\ast) = 0$, but
instead, it is given by
\begin{equation}
  c_i(\vb*{x}^\ast) \approx - \frac{\lambda_i^\ast}{\mu}
\end{equation}
for some fixed parameter $\lambda_i^\ast$.\footnote{$\vb*{\lambda}^\ast$
  is the Lagrange multiplier vector satisfying the first-order
  necessary conditions for optimality, also known as
  Karush--Kuhn--Tucker conditions. See Eq.~(\ref{eq:KKT}).}
The feasibility conditions are satisfied in the limit of $\mu \to
\infty$, but then the penalty function will be dominated by huge
penalty terms that may lead to inaccuracy in the numerical
calculation.
Therefore, we need an algorithm where the solutions more nearly
satisfy the constraints even for moderate values of $\mu$.

The AUGLAG method extends the quadratic penalty method by adding
Lagrangian multipliers to the objective function,
\begin{equation}
  \mathcal{L}_A(\vb*{x}, \vb*{\lambda};\, \mu)
  = f(\vb*{x}) - \sum_i \lambda_i c_i(\vb*{x}) + \frac{\mu}{2} \sum_i
  c_i^2 (\vb*{x}) .
\end{equation}
The Lagrangian multiplier vector $\vb*{\lambda}$ is an input at each
iteration step, not to be determined as in the conventional Lagrangian
multiplier method. For a point $\vb*{x}_k$, the
feasibility condition is now given by
\begin{equation}
  c_i (\vb*{x}_k) \approx - \frac{1}{\mu_k} \left( \lambda_i^\ast -
    \lambda_i^k \right),
\end{equation}
thus, the infeasibility will be much smaller than $1 / \mu_k$ for
$\lambda_i^k \to \lambda_i^\ast$. Namely, the convergence of the AUGLAG
method can be assured without taking $\mu$ to be increasing
indefinitely.
At the $k$th step of iterations, the Lagrangian multiplier vector is
updated as
\begin{equation}
  \lambda_i^k = \lambda_i^{k-1} - \mu_{k-1} c_i (\vb*{x}_{k-1})
  \label{eq:auglang_lambda}
\end{equation}
using the approximation solution $\vb*{x}_{k-1}$ at the previous step.

The most popular software package for the practical AUGLAG method is
\texttt{LANCELOT}~\cite{doi:10.1137/0728030,
  doi:10.1137/1.9780898719857}, and the implementation of
\texttt{OPTIMASS} is inspired in part by the package.
Note that, in each iteration, unconstrained minimization is performed
for given values of $\vb*{\lambda}$ and $\mu$ updated by the AUGLAG
algorithm.
In \texttt{OPTIMASS}, a combination of Migrad and Simplex algorithms
is adopted for the unconstrained minimization: Simplex finds a local
solution at first, and then Migrad makes use of the solution as an
initial guess for searching a minimum.

We have briefly looked over the AUGLAG method, which is the chosen
algorithm of \texttt{OPTIMASS} for calculating $M_2$.
However, we note that the AUGLAG method is not the only available way
for performing constrained minimization.
There are other well-known methods such as the sequential quadratic
programming (SQP)~\cite{Wilson:1963, Palomares:1974, Han:1976,
  Han:1977, Powell:1978} and the interior-point
method~\cite{Forsgren:2002, gould_orban_toint_2005} that can be applied to the
same problem.
In Ref.~\cite{Cho:2015laa}, it was shown that the implementation using
the AUGLAG method effectively achieved the minimization well.
But, it did not mention the particular reason for the choice of the
algorithm, nor show benchmark studies for comparisons to other
methods.
This motivates us to study other numerical methods for calculating the
$M_2$ variables.

For comparison with the AUGLAG method, we employ the SQP
method. We have chosen the SQP method since it is known to be the most
efficient unless the number of free variables is too large, and it
requires fewer function evaluations in comparison with AUGLAG
methods~\cite{Nocedal:2006}. We will compare their performance in the
next section.

The SQP method can be derived by applying Newton's method for solving
nonlinear equations to the condition for the local solution of
constrained minimization problems.
Here we closely follow the description of the SQP method given in
Ref.~\cite{Nocedal:2006}.
Considering only equality constraints, we define the Lagrangian
function for the problem (\ref{eq:const_min}) as
\begin{equation}
  \mathcal{L} (\vb*{x}, \, \vb*{\lambda}) = f(\vb*{x}) - \sum_i
  \lambda_i c_i (\vb*{x}) .
\end{equation}
For continuously differentiable functions $f$ and $c_i$, and a local
solution $\vb*{x}^\ast$ to the problem, there exists a Lagrangian
multiplier vector $\vb*{\lambda}^\ast$ such that the following
conditions are satisfied:
\begin{align}
  \grad_{\vb*{x}} \mathcal{L} (\vb*{x}^\ast, \vb*{\lambda}^\ast)
  &= \vb*{0}, \nonumber\\
  c_i (\vb*{x}^\ast)
  &= 0 . 
    \label{eq:KKT}
\end{align}
They are known as the first-order necessary conditions for
optimality, or the Karush--Kuhn--Tucker (KKT) conditions.

Newton's method is an algorithm widely used for finding the roots of
real-valued equations.
It successively improves the approximation to the roots using the
first derivatives of the equations.
For a continuously differentiable vector function $F(\vb*{x})$:
$\mathbb{R}^n \to \mathbb{R}^n$, the iteration is given by
\begin{equation}
  \vb*{x}_{k + 1} = \vb*{x}_k + \vb*{p}_k ,
\end{equation}
where the search direction $\vb*{p}_k$ is the solution of
\begin{equation}
  J( \vb*{x}_k) \vb*{p}_k = - F(\vb*{x}_k).
\end{equation}
Here $J (\vb*{x})$ is the Jacobian matrix of $F (\vb*{x})$, $J
(\vb*{x}) = \grad F(\vb*{x})$.

We now consider an equality-constrained minimization problem,
\begin{equation}
  \min_{\vb*{x} \in \mathbb{R}^n} f(\vb*{x})
  \quad \text{subject to}\,\,\, c (\vb*{x}) = \vb*{0} ,
  \label{eq:eq_const_min}
\end{equation}
where $f: \mathbb{R}^n \to \mathbb{R}$ and $c: \mathbb{R}^n \to
\mathbb{R}^m$.
The KKT conditions~(\ref{eq:KKT}) of the problem can be written as
\begin{equation}
  F(\vb*{x}, \, \vb*{\lambda}) =
  \begin{pmatrix}
    \grad{f(\vb*{x})} - A (\vb*{x})^\mathsf{T} \vb*{\lambda} \\
    c (\vb*{x})
  \end{pmatrix} = \vb*{0}. \label{eq:KKT_F}
\end{equation}
It has $n + m$ unknown parameters, $\vb*{x}$ and $\vb*{\lambda}$.
$A$ is an $m \times n$ matrix, the Jacobian of the constraints:
\begin{equation}
  A (\vb*{x}) = \left( \grad c_1 (\vb*{x}),\, \dots, \, \grad c_m
    (\vb*{x}) \right)^\mathsf{T} .
\end{equation}
And, the Jacobian matrix of $F (\vb*{x}, \, \vb*{\lambda})$ is given
by
\begin{equation}
  J (\vb*{x}, \, \vb*{\lambda}) =
  \begin{pmatrix}
    \grad_{\vb*{x} \vb*{x}}^2 \mathcal{L}(\vb*{x}, \, \vb*{\lambda})
    & - A (\vb*{x})^\mathsf{T} \\
    A (\vb*{x}) & \vb*{0}
  \end{pmatrix} .
\end{equation}
For the iterate $(\vb*{x}_k$, $\vb*{\lambda}_k)$,
applying Newton's method to the KKT conditions~(\ref{eq:KKT_F}) gives
us the next iterate,
\begin{equation}
  \begin{pmatrix}
    \vb*{x}_{k + 1} \\
    \vb*{\lambda}_{k + 1}
  \end{pmatrix} =
  \begin{pmatrix}
    \vb*{x}_{k} \\
    \vb*{\lambda}_{k}
  \end{pmatrix} +
  \begin{pmatrix}
    \vb*{p}_{x} \\
    \vb*{p}_{\lambda}
  \end{pmatrix} ,
\end{equation}
where $\vb*{p}_{x}$ and $\vb*{p}_{\lambda}$ are the solutions of
\begin{equation}
  \begin{pmatrix}
    \grad_{\vb*{x} \vb*{x}}^2 \mathcal{L}(\vb*{x}, \, \vb*{\lambda})
    & - A (\vb*{x})^\mathsf{T} \\
    A (\vb*{x}) & \vb*{0}
  \end{pmatrix}
  \begin{pmatrix}
    \vb*{p}_{x} \\
    \vb*{p}_{\lambda}
  \end{pmatrix} = -
  \begin{pmatrix}
    \grad{f(\vb*{x})} - \grad A (\vb*{x})^\mathsf{T} \vb*{\lambda} \\
    c (\vb*{x})
  \end{pmatrix} .
\end{equation}
It turns out that the iterate generated by the application of Newton's
method is equivalent to modeling the problem~(\ref{eq:eq_const_min})
using the quadratic subproblem at the iterate
$(\vb*{x}_k$, $\vb*{\lambda}_k)$,
\begin{align}
  & \min_{\vb*{p} \in \mathbb{R}^n} \left[ \mathcal{L}(\vb*{x}_k, \,
    \vb*{\lambda}_k) +  \grad_{\vb*{x}} \mathcal{L}
    (\vb*{x}_k, \, \vb*{\lambda}_k)^\mathsf{T}
    \vb*{p} + \frac{1}{2} \vb*{p}^\mathsf{T} \grad_{\vb*{x} \vb*{x}}^2
    \mathcal{L} (\vb*{x}_k, \, \vb*{\lambda}_k) \vb*{p}
    \right] \nonumber\\
  & \text{subject to}\,\,\, A (\vb*{x}_k) \, \vb*{p} +
    c (\vb*{x}_k) = \vb*{0}.
\end{align}
The objective function in the above is a quadratic approximation
of the Lagrangian function. Therefore, it can be argued that we have
replaced the constrained minimization problem~(\ref{eq:eq_const_min})
by the problem of minimizing the {\em quadratic} approximation of the
Lagrangian function subject to the {\em linear} approximation of the
constraints.
This is the SQP framework, which enables us to derive the SQP
algorithm for nonlinearly constrained minimization problems.
The SQP framework can easily be extended to optimization problems with
inequality constraints.

The SQP framework has a unique solution $(\vb*{p}_k$,
$\vb*{\lambda}_k^\prime)$ satisfying
\begin{align}
  \grad_{\vb*{x} \vb*{x}}^2 \mathcal{L} (\vb*{x}_k, \,
  \vb*{\lambda}_k) \vb*{p}_k + \grad f(\vb*{x}_k) -
  A (\vb*{x}_k)^\mathsf{T} \vb*{\lambda}_k^\prime
  &= \vb*{0}, \nonumber\\
  A (\vb*{x}_k) \, \vb*{p}_k + c (\vb*{x}_k)
  &= \vb*{0} ,
    \label{eq:sqp_framework}
\end{align}
if the following assumptions hold:
\begin{enumerate}
\item[(a)] The Jacobian matrix of the constraints $A(\vb*{x})$ has
  full row rank, {\em i.e.}, the constraint gradients are linearly
  independent.
\item[(b)] The Hessian matrix of the Lagrangian function
  $\grad_{\vb*{x} \vb*{x}}^2 \mathcal{L} (\vb*{x}, \, \vb*{\lambda})$
  is positive definite on the tangent space of the constraints,
  $\vb*{d}^\mathsf{T} \grad_{\vb*{x} \vb*{x}}^2 \mathcal{L} \, \vb*{d}
  > 0$ for all $\vb*{d} \neq \vb*{0}$ such that $A(\vb*{x}) \vb*{d} =
  \vb*{0}$.
\end{enumerate}
After solving the equations, the new iterate $(\vb*{x}_{k + 1}$,
$\vb*{\lambda}_{k + 1})$ are given by
\begin{equation}
  \vb*{x}_{k + 1} = \vb*{x}_k + \vb*{p}_k, \quad
  \vb*{\lambda}_{k + 1} = \vb*{\lambda}_k^\prime .
\end{equation}

As can be seen in (\ref{eq:sqp_framework}),
the SQP method effectively uses the first and second-order derivative
information of the objective and constraint functions.
On the other hand, in the case of the AUGLAG method, the derivatives
are used only in the convergence test except for the minimization to
solve the subproblem.
As long as the derivatives are well-defined over the feasible set, the
SQP method performs very efficiently to find the local minimum.
Moreover, it provides the next iterate $\vb*{x}_{k+1}$ as
well as $\lambda_{k+1}$ based on the current estimate of the local
solution on top of the one found by sub-algorithm such as the
Broyden–Fletcher–Goldfarb–Shanno (BFGS) algorithm, which
will briefly be shown shortly.
Meanwhile, the AUGLAG method only provides
$\vb*{\lambda}_{k+1}$ as in Eq.~(\ref{eq:auglang_lambda}), while
$\vb*{x}_{k+1}$ is determined solely by the sub-algorithm at each
iterate.
Therefore, the performance of the AUGLAG method can also depend highly
on the choice of the sub-algorithm.

The gradients of functions in the $M_2$ variables can be
analytically obtained event by event.
To present the gradients, we define invariant mass functions as
\begin{align}
  M_{(i)}^2 \equiv (P_i + k_i)^2 .
\end{align}
The gradients of the mass functions at $\vb*{k} =
(k_{1x}, \, k_{1y}, \, k_{1z}, \, k_{2z})$ are given by
\begin{align}
  \grad_{\vb*{k}} M_{(1)} (P_1, \, k_1)
  &=
  \frac{1}{M_{(1)} e_1 }
  \begin{pmatrix}
    E_1 k_{1x} - e_1 P_{1x} \\
    E_1 k_{1y} - e_1 P_{1y} \\
    E_1 k_{1z} - e_1 P_{1z} \\
    0
  \end{pmatrix}, \nonumber\\
  \grad_{\vb*{k}} M_{(2)} (P_2, \, k_2)
  &=
  \frac{1}{M_{(2)} e_2 }
  \begin{pmatrix}
    E_2 (k_{1x} - \slashed{P}_x) + e_2 P_{2x} \\
    E_2 (k_{1y} - \slashed{P}_y) + e_2 P_{2y} \\
    0 \\
    E_2 k_{2z} - e_2 P_{2z}
  \end{pmatrix} .
  \label{eq:grad_mass}
\end{align}
Using the invariant mass functions, the gradient of the objective function of
the $M_2$ variables~(\ref{eq:M2_obj}) is written as follows:
\begin{equation}
  \grad_{\vb*{k}} f (\vb*{k}) =
  \begin{cases}
    \grad_{\vb*{k}} M_{(1)} (p_1, \, k_1)
    & \text{if } M_{(1)} (p_1, \, k_1) \geq M_{(2)} (p_2, \, k_2) , \\
    \grad_{\vb*{k}} M_{(2)} (p_2, \, k_2)
    & \text{if } M_{(1)} (p_1, \, k_1) < M_{(2)} (p_2, \, k_2) .
  \end{cases}
\end{equation}
Furthermore, the on-shell mass constraints can be expressed by
\begin{align}
  c_A &= M_{(1)} (p_1, \, k_1) - M_{(2)} (p_2, \, k_2) , \nonumber\\
  c_B &= M_{(1)} (q_1, \, k_1) - M_{(2)} (q_2, \, k_2) ,
\end{align}
for $A_i$ and $B_i$, respectively.
Therefore, it is straightforward to obtain the gradients of the
constraints by using $\grad_{\vb*{k}} M_{(i)}$ in~(\ref{eq:grad_mass})
as well,
\begin{align}
  \grad_{\vb*{k}} c_A
  &= \grad_{\vb*{k}} M_{(1)} (p_1, \, k_1) -
  \grad_{\vb*{k}} M_{(2)} (p_2, \, k_2) , \nonumber\\
  \grad_{\vb*{k}} c_B
  &= \grad_{\vb*{k}} M_{(1)} (q_1, \, k_1) -
  \grad_{\vb*{k}} M_{(2)} (q_2, \, k_2) .
\end{align}

Meanwhile, the exact form of the Hessian matrix of the Lagrangian
$\grad_{\vb*{x} \vb*{x}}^2 \mathcal{L}$ is not necessary if we employ
quasi-Newton approximation, where only the gradient information is
required.
In the quasi-Newton method, the approximation of the Hessian matrix
$B_k$ satisfies the so-called secant condition,
\begin{equation}
  B_{k + 1} (\vb*{x}_{k + 1} - \vb*{x}_k) =
  \grad \mathcal{L} (\vb*{x}_{k+1} , \, \vb*{\lambda}_{k + 1})
  - \grad \mathcal{L} (\vb*{x}_{k} , \, \vb*{\lambda}_{k}) .
\end{equation}
The strategy is that we compute $(\vb*{x}_{k + 1}$, $\vb*{\lambda}_{k
  + 1})$ for a given $B_k$, and then update $B_k$ as
\begin{equation}
  B_{k + 1} = B_k + U_k.
\end{equation}
To determine $U_k$ uniquely, it is necessary to impose additional
conditions, which differ by algorithms.
The most popular and powerful method is the BFGS
algorithm, and we will use the
algorithm in our implementation for the $M_2$ variables.

\section{\label{sec:performance}Implementation and performance comparisons}

\noindent
We are now in a position to describe our implementation for
calculating the $M_2$ variables.
For employing the numerical minimization algorithms, we use
\texttt{NLopt}, a publicly available software library for nonlinear
optimization~\cite{cite:nlopt}.
Since \texttt{NLopt} includes several numerical algorithms with a uniform
interface, we have also tested the other algorithms not examined
in the previous section.
Among them, we find that the SQP and AUGLAG methods with the BFGS
update perform the best in terms of accuracy and speed for calculating $M_2$.
In \texttt{NLopt}, the SQP algorithm is based on the implementation
described in Refs.~\cite{kraft1988software, Kraft:1994}, and the
implementation of the AUGLAG algorithm follows
Refs.~\cite{doi:10.1137/0728030, Birgin:2008}.
Our studies on the numerical algorithms have brought yet another
library for the $M_2$ variables, which we dub
\texttt{YAM2}~\cite{Park:YAM2}.
The instructions for installation and usage of \texttt{YAM2} are given
in Sec.~\ref{sec:install}.
For comparison, we have also added an interface for using the
AUGLAG method with the Nelder-Mead Simplex algorithm to the library.

Currently, the set of the algorithms employed in \texttt{YAM2} is as follows:
\begin{itemize}
\item \texttt{SQP} $+$ \texttt{BFGS}: SQP algorithm with the BFGS
  update,
\item \texttt{AUGLAG} $+$ \texttt{BFGS}: AUGLAG algorithm with the
  BFGS update,
\item \texttt{AUGLAG} $+$ \texttt{Simplex}: AUGLAG algorithm with the
  Nelder-Mead Simplex method,
\item a combination of the above algorithms.
\end{itemize}
In the combination of the algorithms, the $M_2$ solution is given by
comparing the minima found by \texttt{SQP} $+$ \texttt{BFGS} and
\texttt{AUGLAG} $+$ \texttt{BFGS}. If both algorithms have failed to
find a minimum, the \texttt{AUGLAG} $+$ \texttt{Simplex} algorithm
is used.
We used the combination of the algorithms for the distributions in
Fig.~\ref{fig:M2}.

We must choose an initial guess for the unknown variables $\vb*{k} =
(k_{1x}$, $k_{1y}$, $k_{1z}$, $k_{2z}$) as an input to the algorithms.
The initial guess is important because the numerical minimization
algorithms perform the best in general if the guess is near the
solution, or at least, is not very far from the solution.
We have tested two
kinds of initial guesses: one is that $k_{1x} = \slashed{P}_{1x}/2$,
$k_{1y} = \slashed{P}_{1y}/2$, and $k_{1z} = k_{2z} = 0$, and the
other is the momentum configuration that minimizes the total invariant
mass of the final state, {\em i.e.},
\begin{equation}
  \argmin_{\vb*{k}_1, \, \vb*{k}_2} (p_1 + p_2 + k_1 + k_2)^2
  \quad
  \text{subject to}\,\,\, \vb*{k}_{1T} + \vb*{k}_{2T} =
  \slashed{\vb*{P}}_T .
\end{equation}
It corresponds to the solution of $\hat s_\text{min}$ in
Ref.~\cite{Konar:2008ei}.
We find that both guesses work well, but the latter is slightly better
to find the minimum. In \texttt{YAM2}, we use the solution of the
total invariant mass given above as the initial guess for all the
numerical algorithms.

One of the most important parameters
that can be adopted by the user input is the error tolerance.
It serves as a stopping criterion for the iterations of the algorithms
as well as a measure of the error relative to the solution.
The tolerance can be set for the absolute or relative values of
objective and constraint functions.
One subtlety for setting the tolerance is that the objective and
constraint functions of the $M_2$ variables are mass-dimensionful
quantities whose scales vary event by event.
In order to remove the scale dependence, we rescale all the
masses and momenta of particles by a scale parameter $s$,
\begin{equation}
  p_i \to s^{-1} p_i , \quad
  q_i \to s^{-1} q_i , \quad
  \slashed{\vb*{P}}_T \to s^{-1} \slashed{\vb*{P}}_T,
  \quad
  M_C \to s^{-1} M_C ,
\end{equation}
so that all the quantities become dimensionless.
After the algorithm has completed minimization,
the solution will be scaled back to have the right mass dimension:
$M_2 \to s M_2$ and
$\widetilde{\vb*{k}} \to s \widetilde{\vb*{k}}$.
In \texttt{YAM2}, we have heuristically taken the scale parameter $s$
for a given event to be
\begin{equation}
  s = 8 \sqrt{E_1^2 + E_2^2 + \norm{\slashed{\vb*{P}}_T}^2 + 2
    M_C^2} ,
\end{equation}
where $E_i^2 = p_i^2 + \norm{\vb*{p}_i}^2$ are the squared energies of
visible particles.
On the rescaled objective and constraint functions, we set the tolerance
conditions as follows:
\begin{itemize}
\item for constraints $c_i (\vb*{x})$,
  \begin{equation}
    \abs{c_i (\vb*{x}_k)} < \varepsilon ,
  \end{equation}
\item for objective function $f(\vb*{x})$,
  \begin{align}
    &~ \abs{f(\vb*{x}_k) - f(\vb*{x}_{k -1})} < \varepsilon \times
      10^{-3} \nonumber\\
    \text{or}
    &~ \abs{f(\vb*{x}_k) - f(\vb*{x}_{k -1})} < \frac{\abs{f(\vb*{x}_k) +
      f(\vb*{x}_{k -1})}}{2} \times \varepsilon \times 10^{-3}
  \end{align}
  at the $k$th step.
\end{itemize}
In \texttt{YAM2}, the default value of the $\varepsilon$ parameter is
$10^{-3}$. Users can feed a different value to that.
On occasion, the numerical algorithms fail to find a minimum. In this
case, we find that loosening the tolerance condition is helpful.
Whenever the algorithm throws failure, we increase the $\varepsilon$
parameter to be ten times larger and then restart the algorithm.

It would also be wise to set a maximal number of iterations in order
to avoid increasing the execution time indefinitely.
The maximal number is set to be 5,000.
However, in our experience of numerical studies, a large number of
iterations is a typical symptom that the algorithm is failing to find
the solution. In this situation, it is encouraged to adjust the
tolerance or initial guess rather than increasing the maximal number
of iterations.

To demonstrate and compare the performance of the algorithms, we
consider the di-leptonic top pair process of~(\ref{eq:ttbar}), which
has the decay topology of~(\ref{eq:pair_cascade}).
We have generated Monte Carlo event samples using \texttt{Pythia
  8}~\cite{Sjostrand:2014zea}, and analyzed the parton-level data. The
center-of-mass energy of proton-proton collision has been set to be
$\sqrt{s} = 13$~TeV. The total number of analyzed event samples is
200k.

\begin{figure}[tb!]
  \centering
    \includegraphics[width=0.24\textwidth]{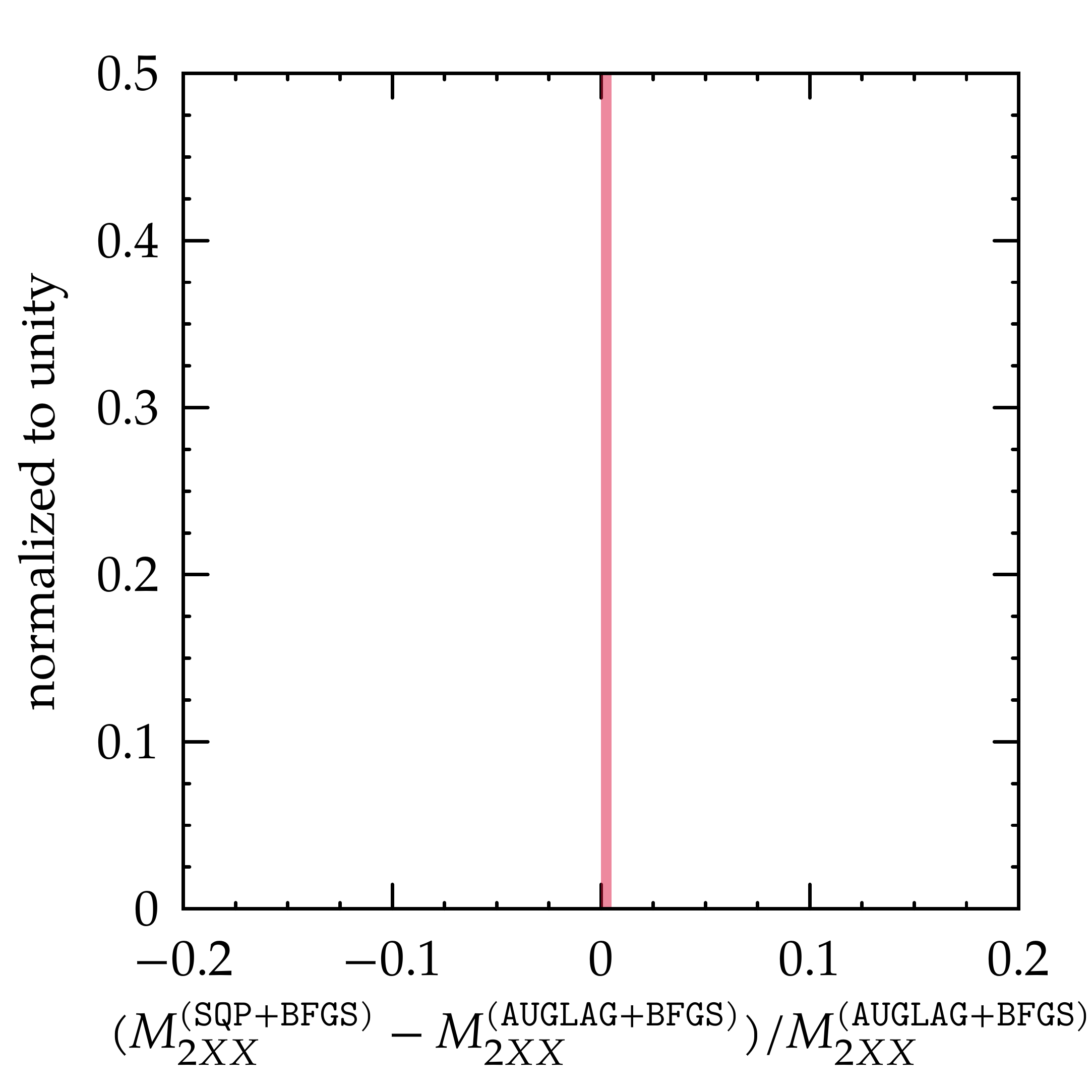}
    \includegraphics[width=0.24\textwidth]{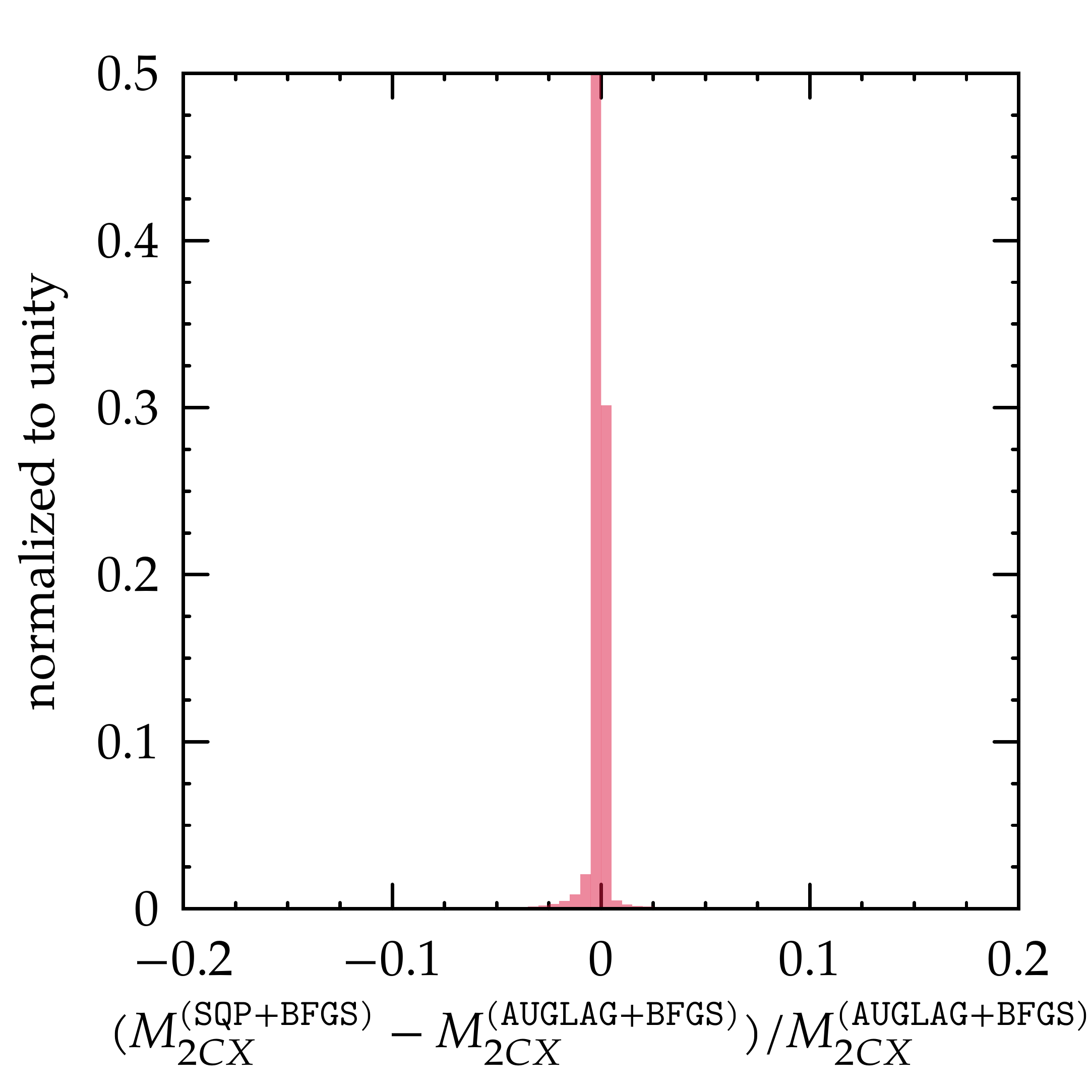}
    \includegraphics[width=0.24\textwidth]{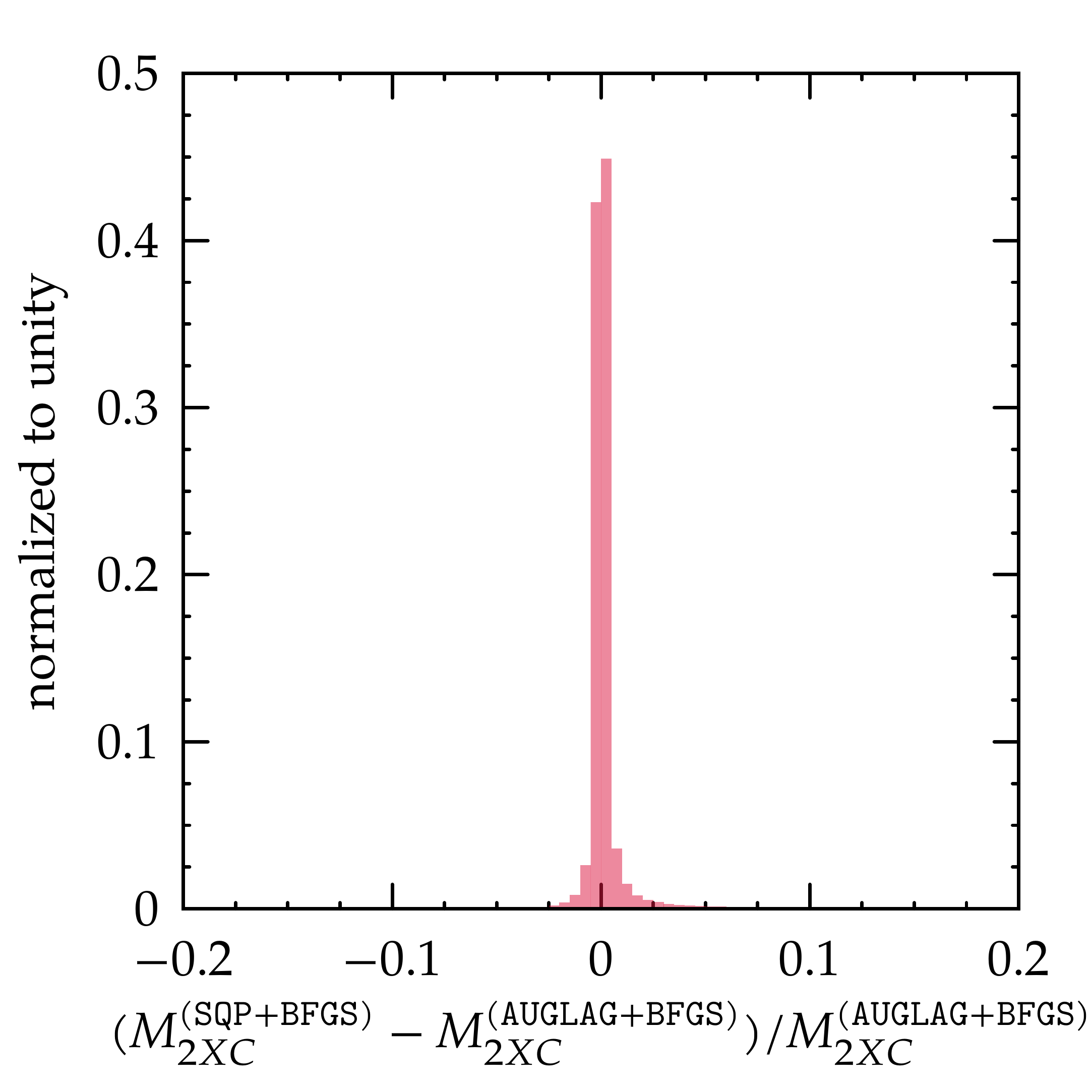}
    \includegraphics[width=0.24\textwidth]{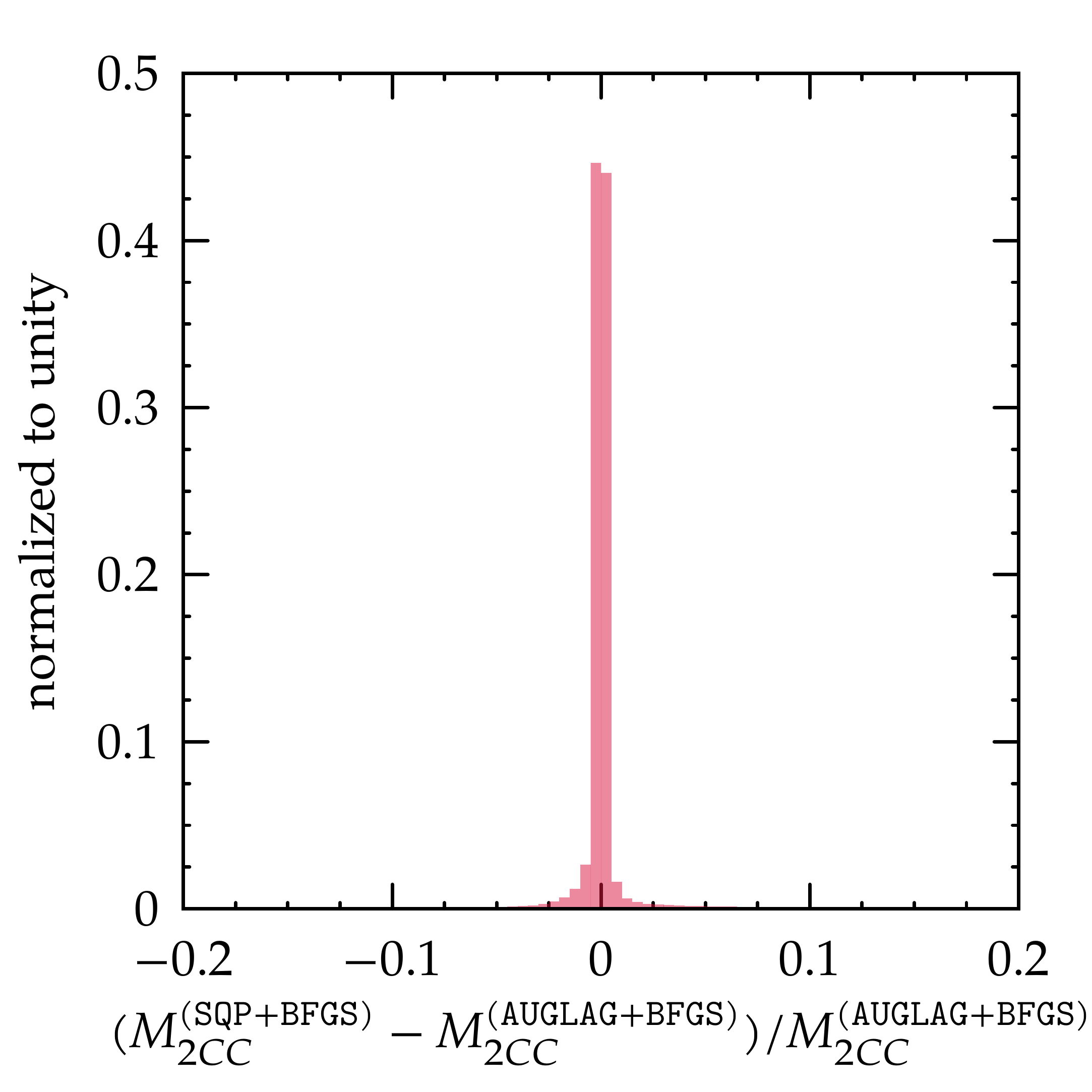}
  \caption{\label{fig:m2_sqp_auglag}
    Histograms for the relative
    differences of the $M_2$ variables calculated
    with SQP and AUGLAG methods. We used the BFGS algorithm for both
    methods.}
\end{figure}

In Fig.~\ref{fig:m2_sqp_auglag}, we show the relative differences
between the $M_2$ variables calculated with the SQP and the AUGLAG
methods.
In the case of $M_{2XX}$, the two methods are the same because there
is no on-shell mass constraints.
In the other cases, the $M_2$ values mostly match
within 0.5\%,
while the SQP method is slightly better for
$M_{2CX}$ and $M_{2CC}$.
Here, saying the better means that the method finds the deeper local
minimum.
We also find that tolerance values smaller than $\varepsilon = 10^{-3}$ do
not improve the result much, so it appears to be an effective choice.
However, it is always worthwhile to check the result by changing the
tolerance before producing the final result in practical physics analyses.

\begin{figure}[tb!]
  \centering
    \includegraphics[width=0.24\textwidth]{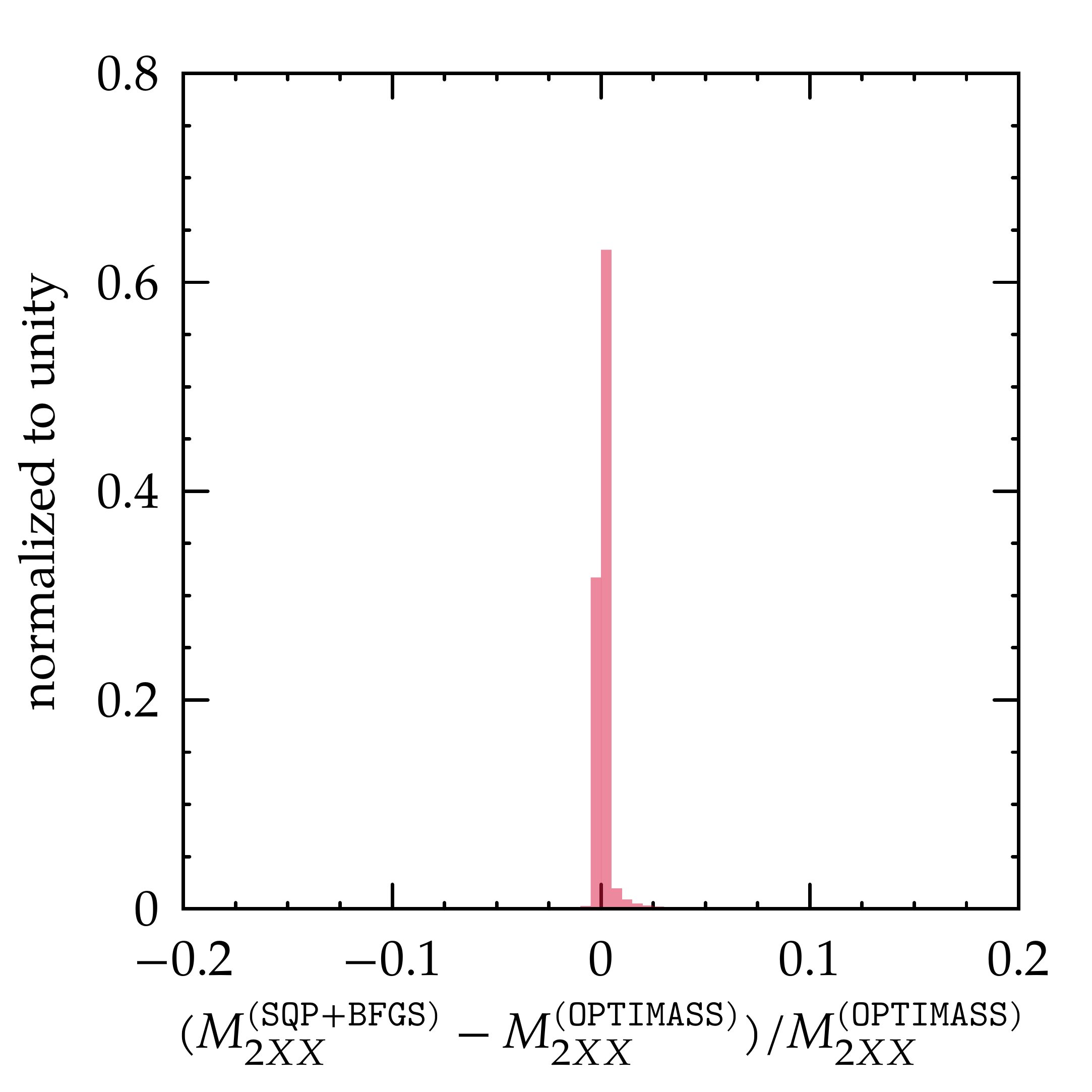}
    \includegraphics[width=0.24\textwidth]{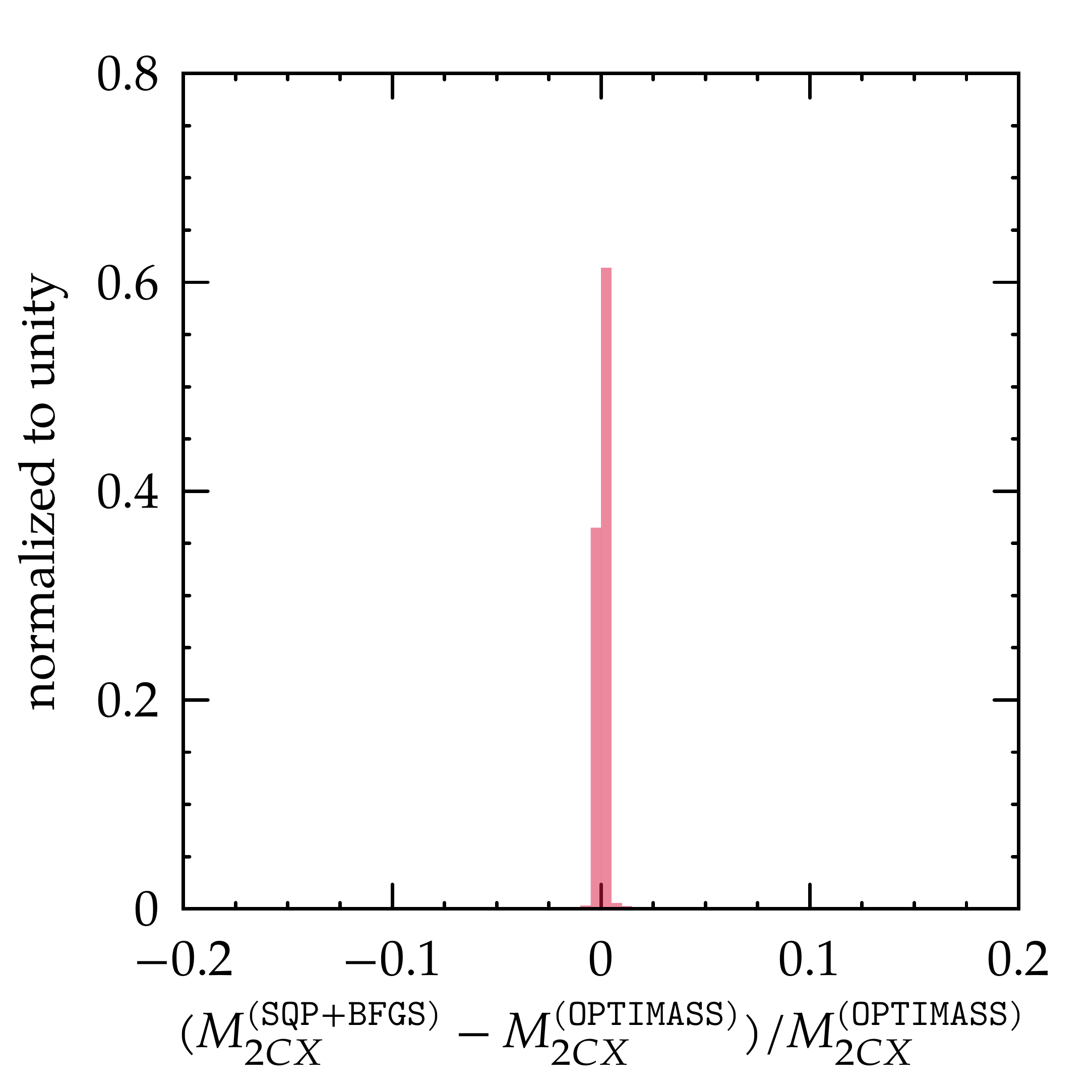}
    \includegraphics[width=0.24\textwidth]{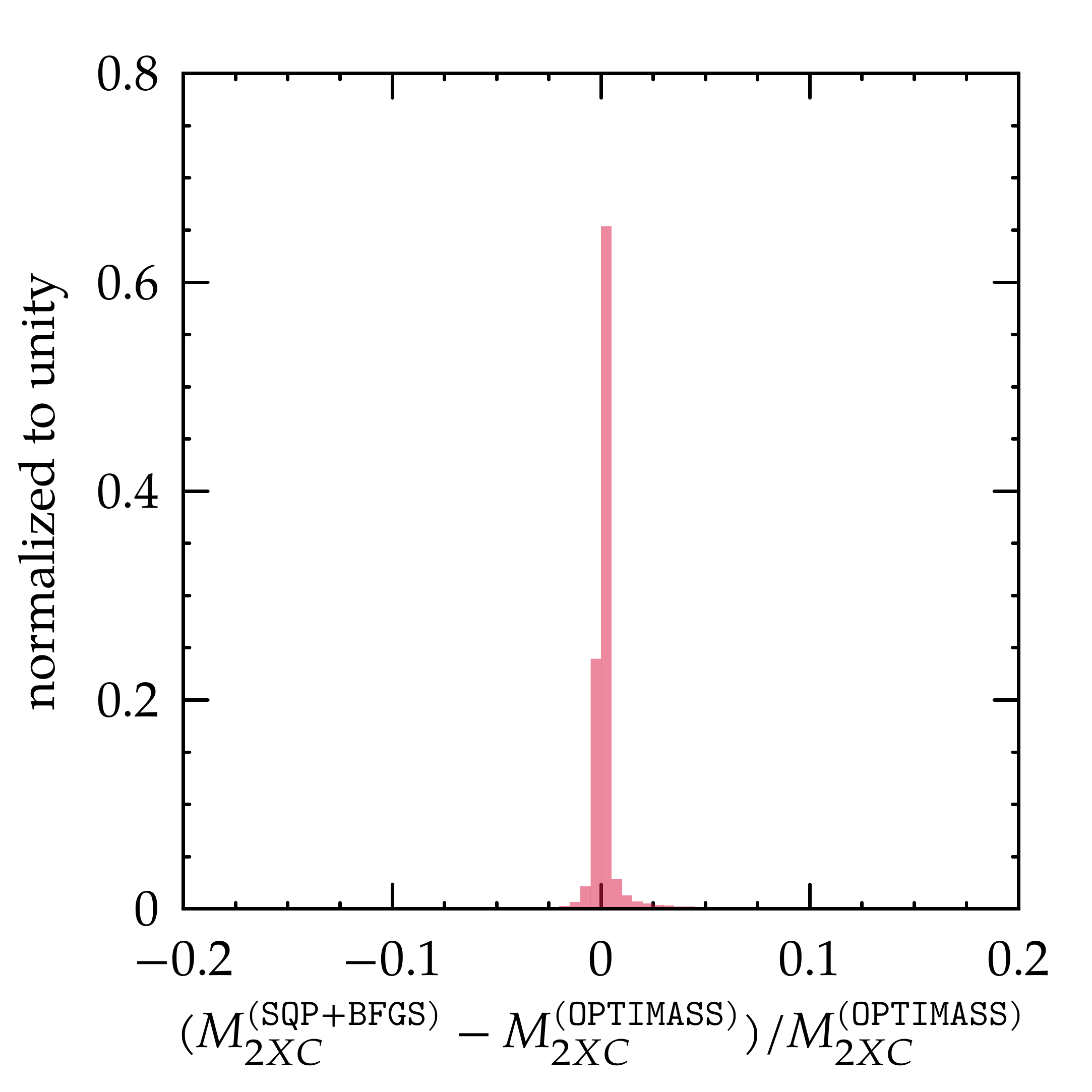}
    \includegraphics[width=0.24\textwidth]{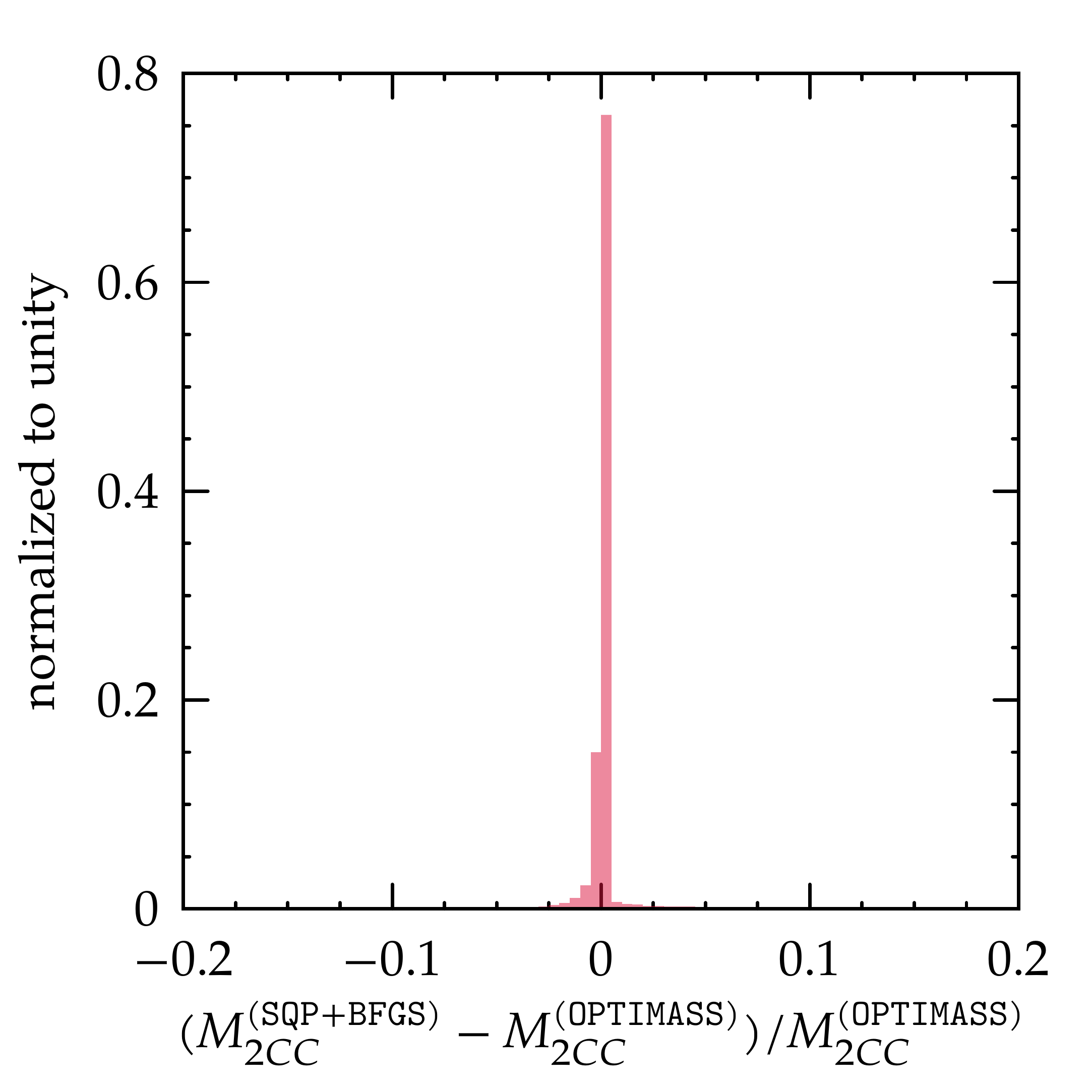}
  \caption{\label{fig:m2_sqp_optimass}
    Histograms for the relative
    differences of the $M_2$ variables calculated
    with the SQP method with BFGS update and \texttt{OPTIMASS}.}
\end{figure}

To compare the result with \texttt{OPTIMASS}, we show the relative
differences between the $M_2$ variables calculated by using the SQP
method and \texttt{OPTIMASS} in Fig.~\ref{fig:m2_sqp_optimass}.
We have used the \texttt{OPTIMASS} version 1.0.3 with the default
setup of parameters included in the package.
The results mostly match each other, and the relative deviations are
only $\lesssim 0.5$\%.
For $M_{2CC}$, the SQP method finds a slightly better
minimum in some events.
Therefore, we find that the numerical methods employed in
\texttt{YAM2} perform well enough.

\begin{figure}[tb!]
  \centering
    \includegraphics[width=0.24\textwidth]{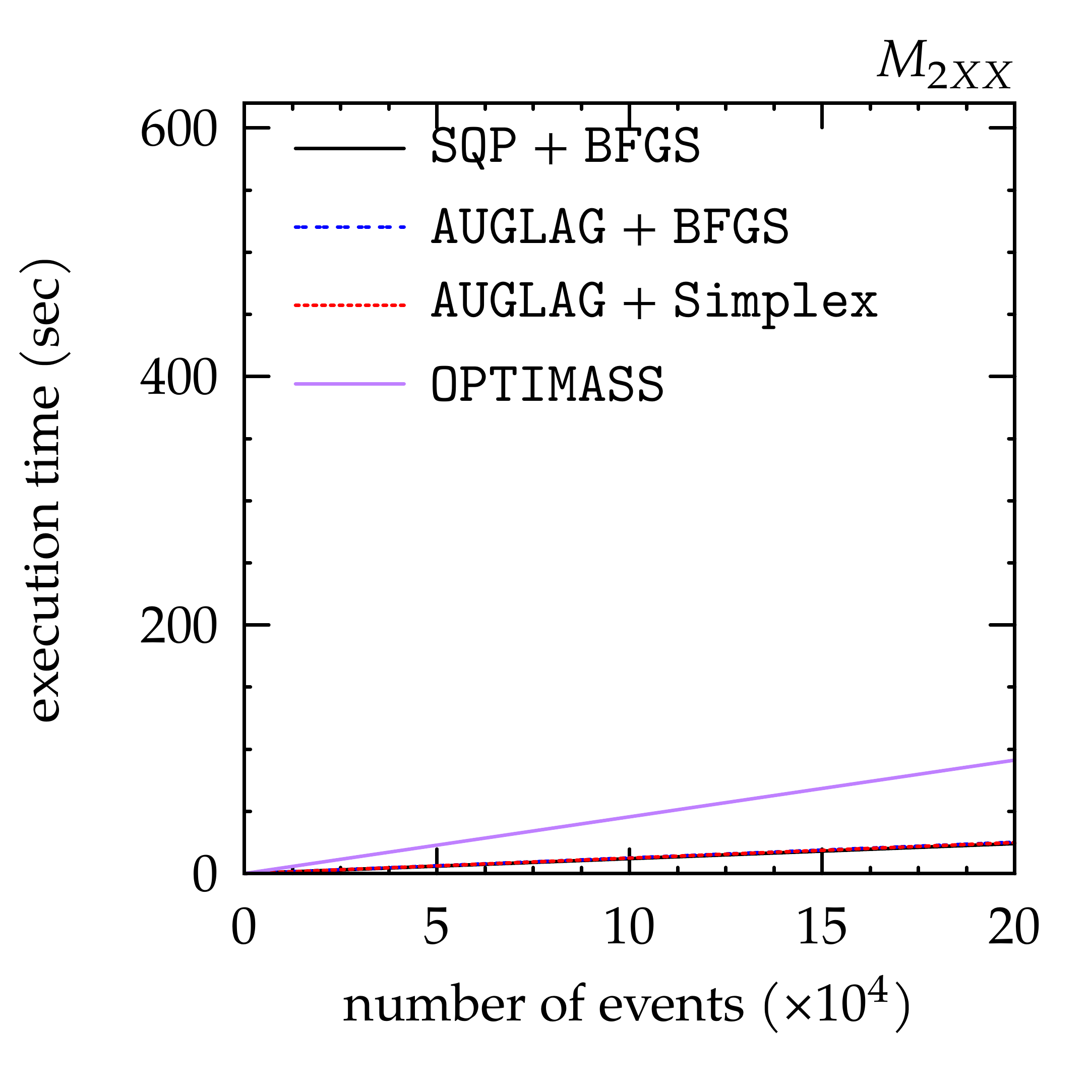}
    \includegraphics[width=0.24\textwidth]{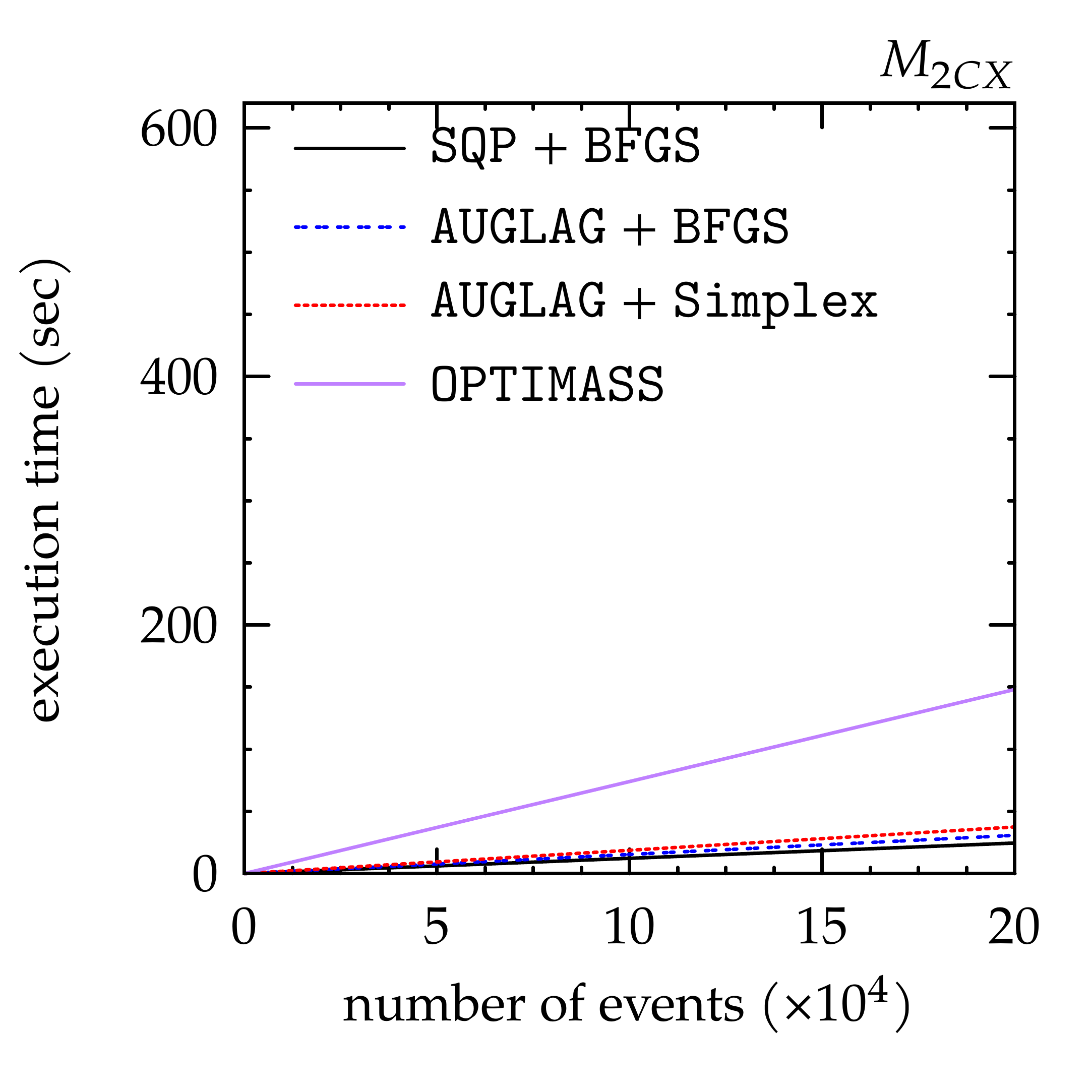}
    \includegraphics[width=0.24\textwidth]{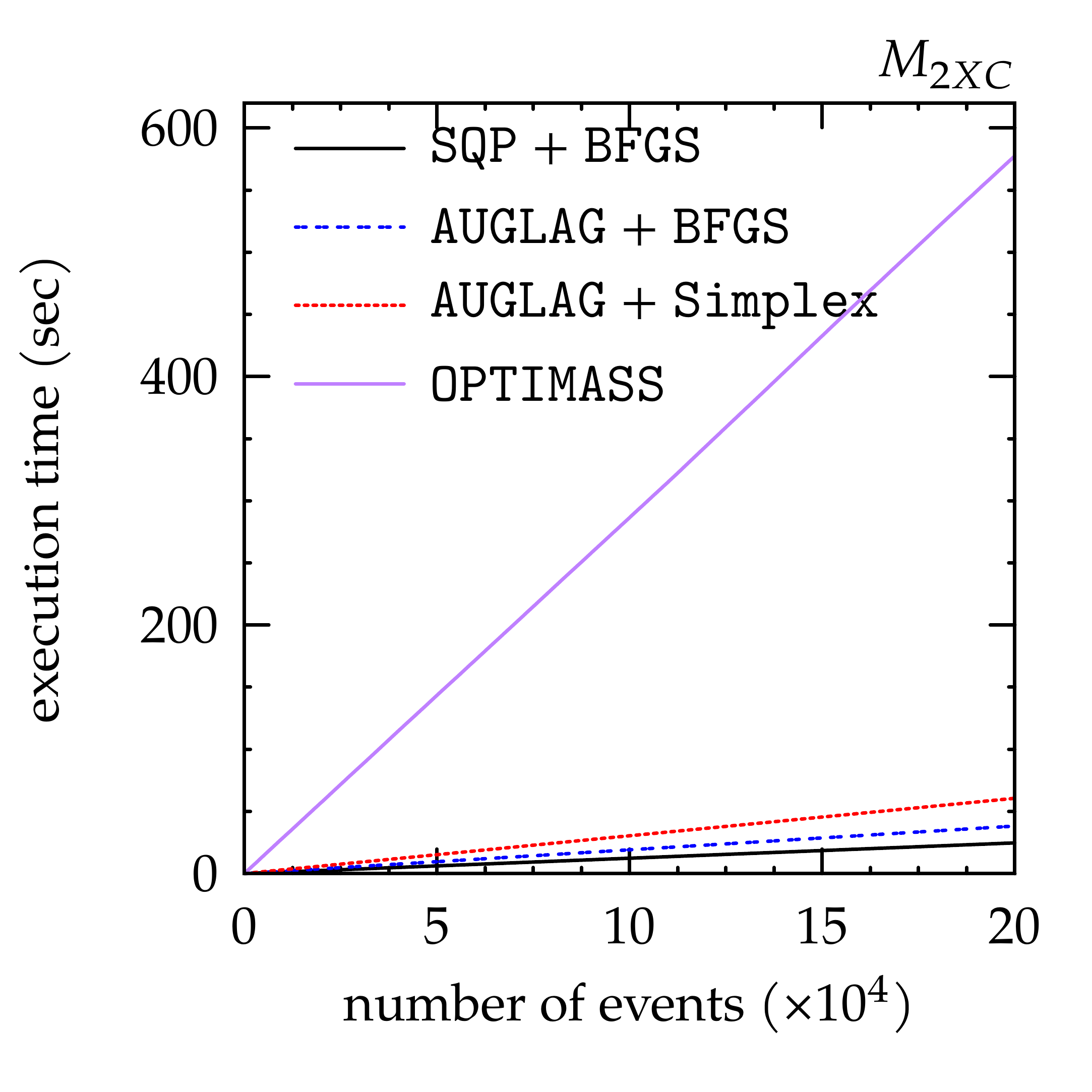}
    \includegraphics[width=0.24\textwidth]{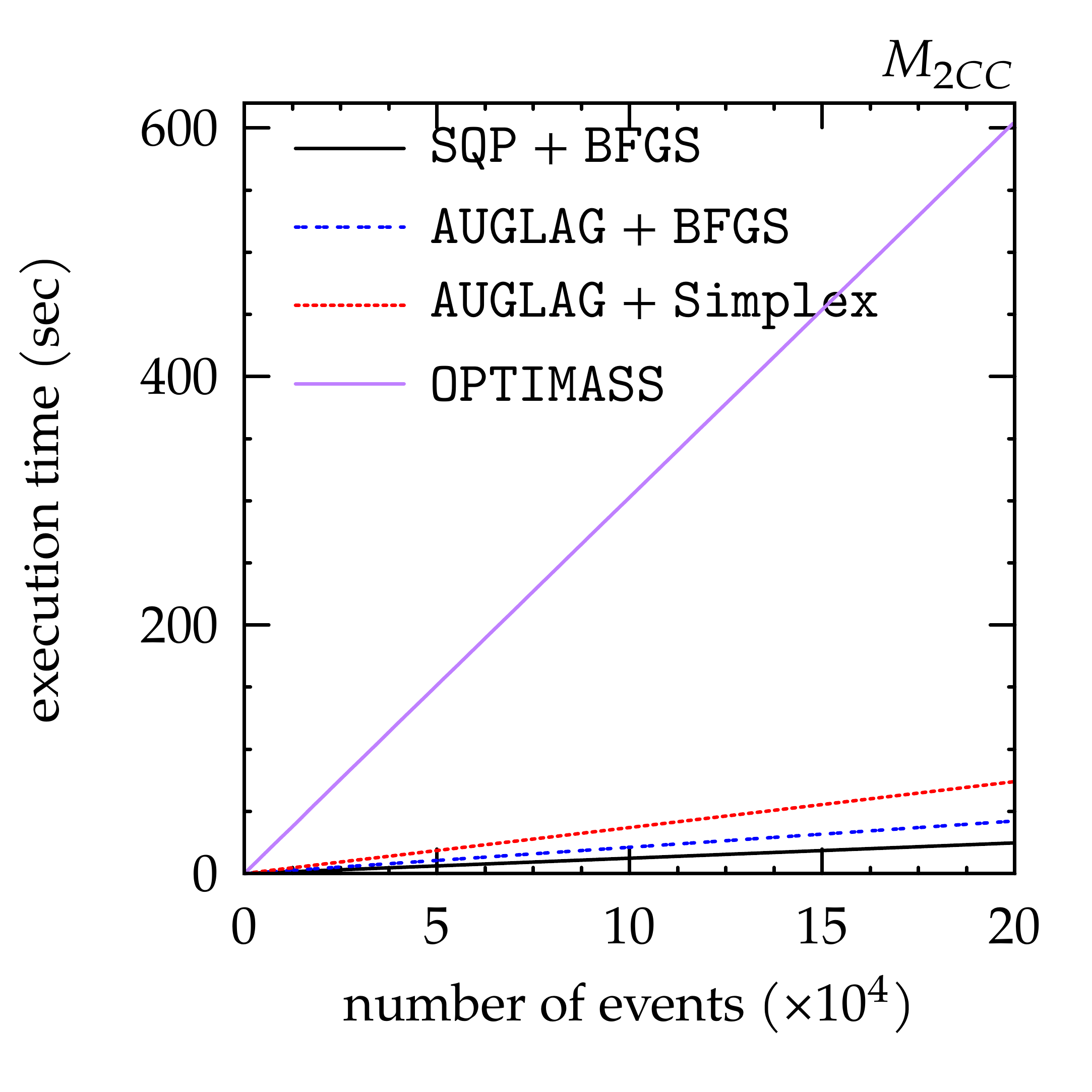}
    \caption{\label{fig:runtime_m2}
      Accumulated execution time of calculating the $M_2$ variables
      using numerical minimization algorithms. The time has been
      measured by using the \texttt{std::chrono} library of
      \protect\CC.}
\end{figure}

We now consider the computational cost of the $M_2$ calculations using
the numerical algorithms. In real situations, we may have to deal with a
tremendous amount of data for physics analyses.
The computational cost is of particular importance, as the integrated
luminosity of the current LHC experiment increases by order of
magnitude, and the High-Luminosity LHC project is on the
horizon~\cite{Apollinari:2017cqg}.
Given limited computing resources and human time, a cost-effective way
without damaging or deteriorating the results will be the most preferable.
Studies of numerical algorithms should be accompanied with a measure
of the execution time.
In Fig.~\ref{fig:runtime_m2}, the accumulated execution time of
calculating the $M_2$ variables is exhibited.
For a fair comparison, we have used the same routines for parsing event
data, and the execution time has been measured by using the
\texttt{std::chrono} library of \CC~in the analysis codes.
All the codes have been compiled and linked by \texttt{g++} of the GNU
Compiler Collection version 10.1 with the optimization level of
\texttt{-O2}.
The resulting executables have been run in a machine with
Intel$^\text{\tiny\textregistered}$
Xeon$^\text{\tiny\textregistered}$
processor E5 3.4~GHz. We did not use run-time parallelization libraries.
For 200k events, the calculation of $M_{2CC}$ using the SQP method
with the BFGS update takes about 25~seconds, while it does about
10~minutes when using \texttt{OPTIMASS}.
Thus, the SQP method implemented in \texttt{YAM2} is faster than
\texttt{OPTIMASS} by $\mathcal{O}(20)$-times.
The combination of the SQP and AUGLAG methods takes about 1~minute,
which is still $\mathcal{O}(10)$-times faster than \texttt{OPTIMASS}.
We have repeated the time measurement a thousand times and found that the
speed upgrade is stable.
Fig.~\ref{fig:runtime_m2} also shows that the SQP method is faster
than the AUGLAG method, and the derivative-dependent algorithm such as
the BFGS is faster than the derivative-free one for all the cases.
We have used the analytic expressions for the gradients given in
Sec.~\ref{sec:algo} for the derivative-dependent
algorithms.\footnote{In \texttt{OPTIMASS}, the gradients are
  calculated numerically by using finite difference method instead of
  the analytic expression.}

\begin{figure}[tb!]
  \centering
    \includegraphics[width=0.24\textwidth]{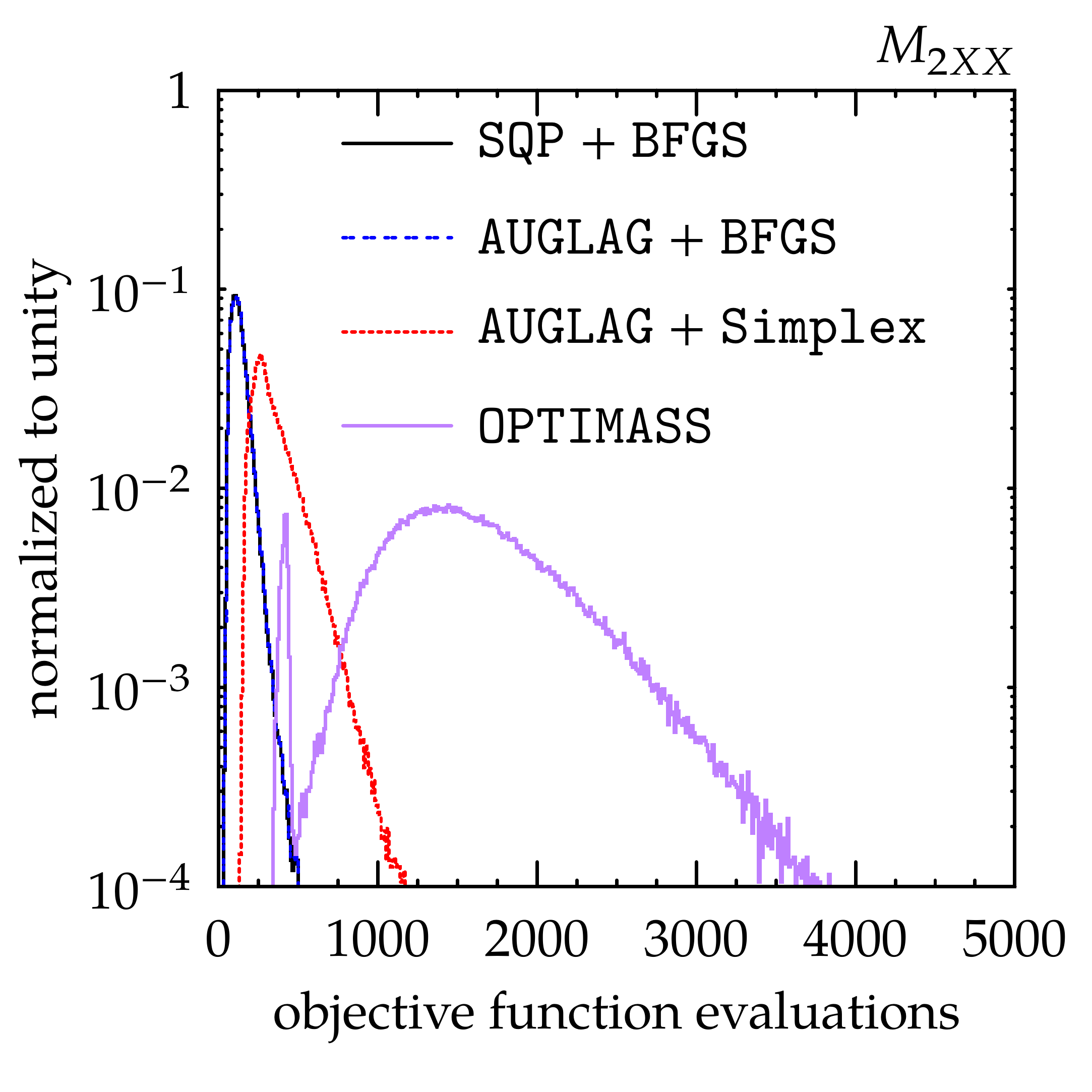}
    \includegraphics[width=0.24\textwidth]{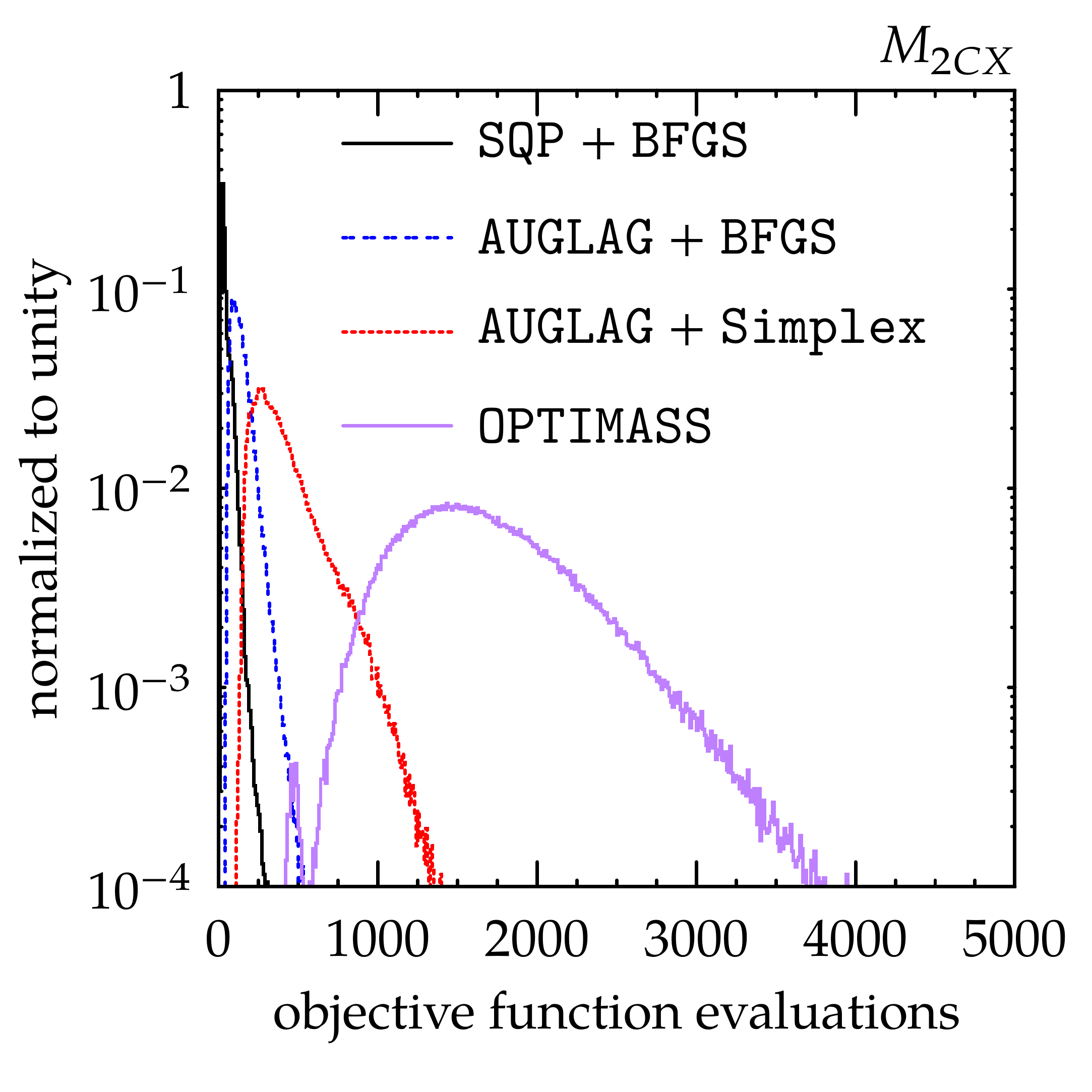}
    \includegraphics[width=0.24\textwidth]{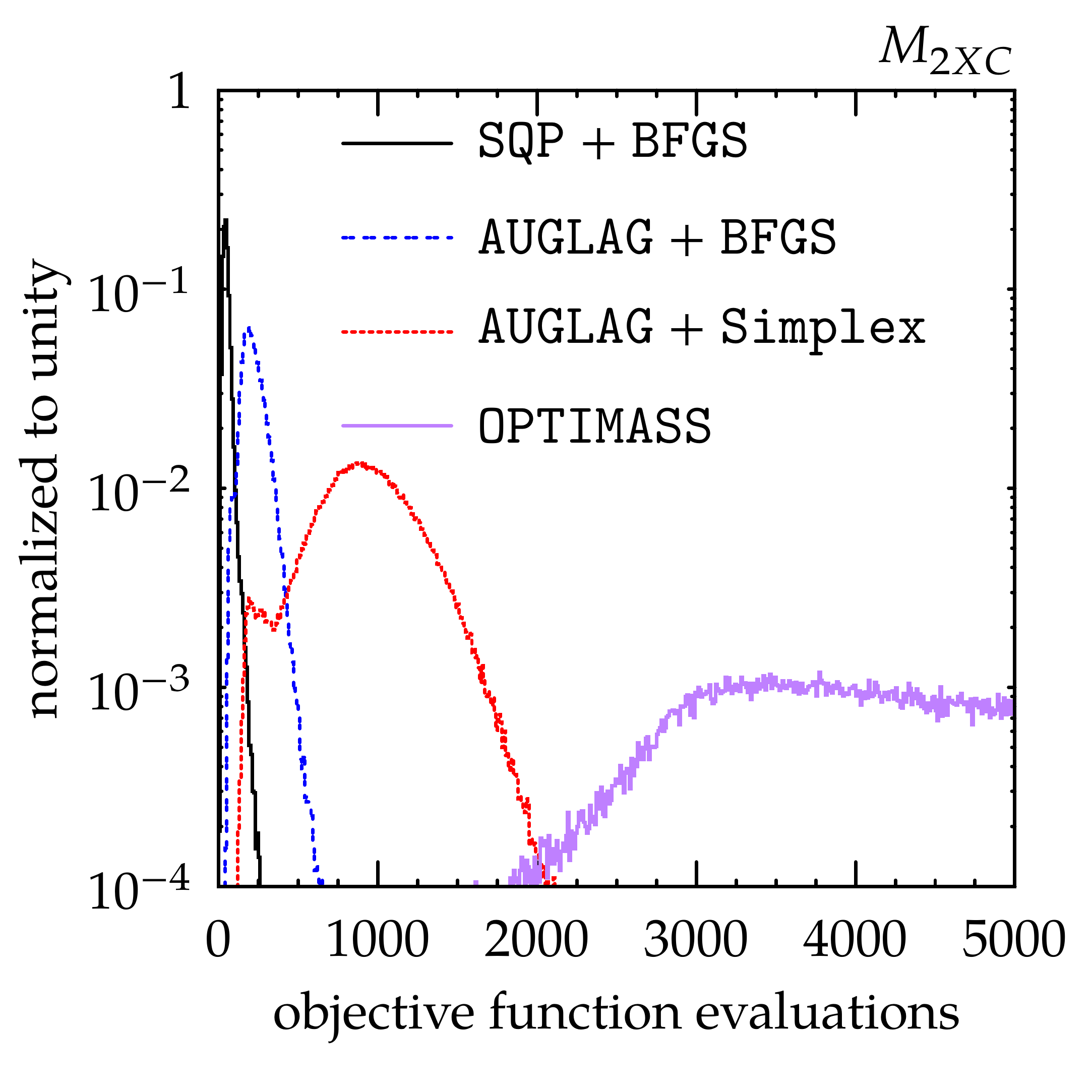}
    \includegraphics[width=0.24\textwidth]{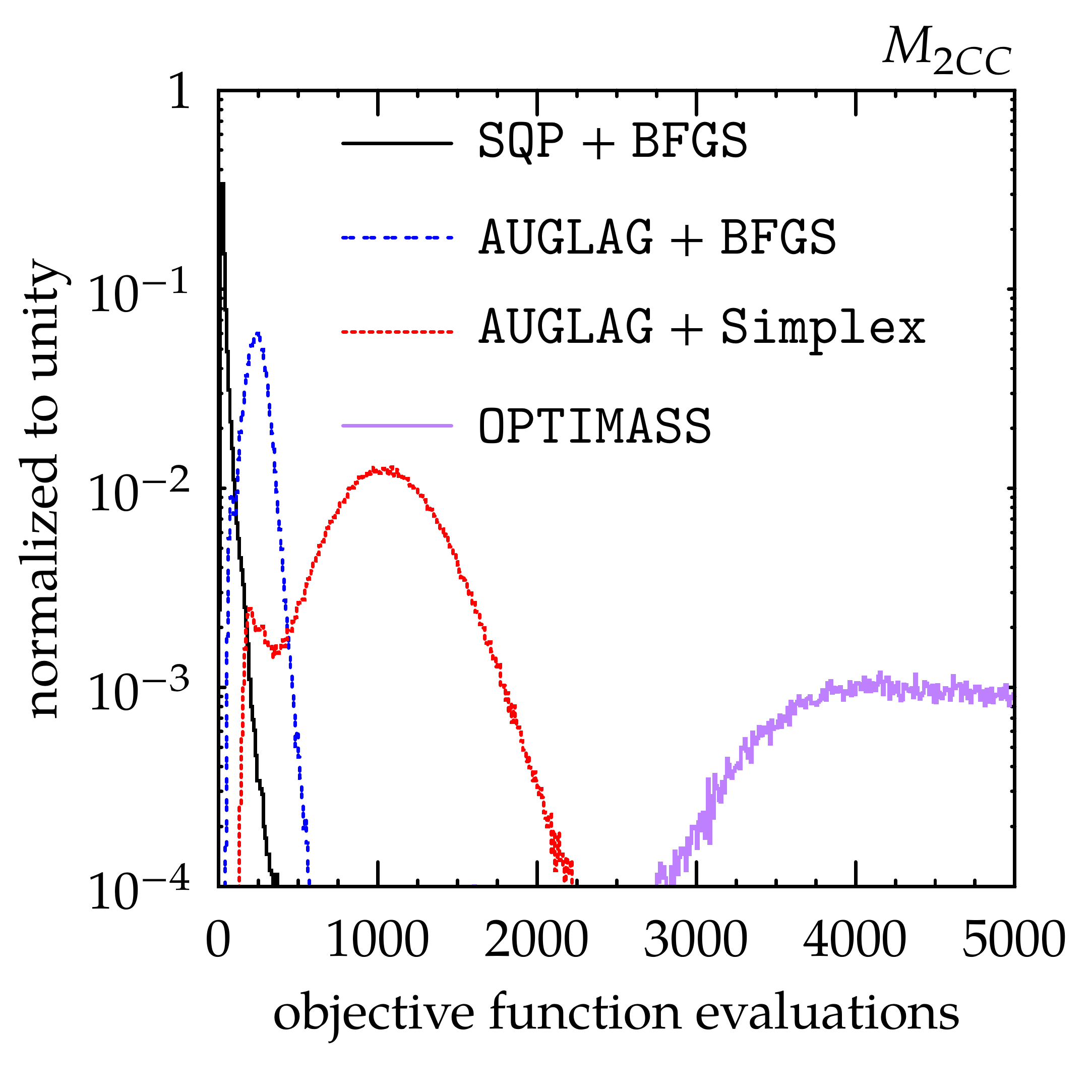}
    \caption{\label{fig:neval_m2}
      Histograms of the number of objective function evaluations for
      calculating the $M_2$ variables using numerical minimization
      algorithms.}
\end{figure}

Another mean for comparing numerical algorithms is to measure the
number of function evaluations. We have counted the number of
objective function evaluations for each event, and the result is shown
in Fig.~\ref{fig:neval_m2}.
We can see that the number for the SQP method is much less than that
of AUGLAG method.
The methods adopting the BFGS update evaluate the objective function
$\mathcal{O}(10)$--$\mathcal{O}(100)$ times per event, while the
number of evaluations is an order of magnitude larger in the case of
the AUGLAG method with the Simplex algorithm.
In \texttt{OPTIMASS}, the number $\gtrsim \mathcal{O}(10^3)$ is much
larger than the others.
We have also measured the amount of system memory used while
executing the analysis codes.
For calculating $M_{2CC}$, the executables from the \texttt{YAM2}
codes consumed about 25~MB memory for processing whole 200k events,
while the \texttt{OPTIMASS} codes consumed about 160~MB\@.
Therefore, we conclude that the implementation of \texttt{YAM2} can
calculate the $M_2$ variables in a much faster and more cost-effective
way.

\section{\label{sec:install}Installation and usage}

\noindent
\texttt{YAM2} is free software under the license specified in the
source code. It is distributed via
\begin{center}
  \url{https://github.com/cbpark/YAM2}\,.
\end{center}
The version of February 16, 2021, is stored in the program library of
Comput.~Phys.~Commun.
Any issue, including bug reports, can be reported through the above
source code repository.
For compiling and linking the source code, it is required to have a
\CC~compiler, supporting the features of the \CC17 revision, and the
\texttt{NLopt} library. Detailed instructions for installing
\texttt{NLopt} by building the source code are given in
Ref.~\cite{NLopt:install}.
In some Linux distributions, it can be installed by using system
package manager.
We have tested our codes with the \texttt{NLopt} version 2.6.2.

The source code of the \texttt{YAM2} library can be built by using the
build automation tool \texttt{make}. If the path to \texttt{NLopt} is
\texttt{/usr/local}, append the path to the \texttt{make} command:
\begin{center}
\begin{varwidth}{\linewidth}
\begin{verbatim}
NLOPT=/usr/local make
\end{verbatim}
\end{varwidth}
\end{center}
The command will build all the source codes, and then generate a static
library file, \texttt{libYAM2.a}, in the \texttt{lib} directory. If the
shared library is necessary, run \texttt{make lib}.
See \texttt{Makefile} for the detail of the compilation flags and path
settings.
The header and library files can also be installed to the other
destination path outside the build directory. If the path to be
installed is \texttt{/usr/local}, run the command as follows.
\begin{center}
\begin{varwidth}{\linewidth}
\begin{verbatim}
DESTDIR=/usr/local make install
\end{verbatim}
\end{varwidth}
\end{center}

The interfaces for using \texttt{YAM2} are defined in the header file
\texttt{yam2.h}.
Users have to add the header to their analysis code through include
directive.
\begin{lstlisting}
#include <yam2.h>
\end{lstlisting}
The type signature of the function for calculating $M_{2CC}$ can be
seen in the following function declaration.
\begin{lstlisting}
std::optional<M2Solution> m2CCSQP(
    const std::optional<InputKinematics> &inp,
    double eps = EPS, int neval = NEVAL);
\end{lstlisting}
The name of the function is descriptive. It will calculate
$M_{2CC}$ using the SQP method. For $M_{2XC}$, the function to use is
\texttt{m2XCSQP}.
The function for calculating $M_{2CC}$
using the AUGLAG method with the BFGS update is
\texttt{m2CCAugLagBFGS}.
The corresponding function using the combination of the SQP and AUGLAG
methods is named \texttt{m2CC}.
In the function declaration given above, one can see that the return
type of the function is \texttt{std::optional} of \texttt{M2Solution}.
The class template \texttt{std::optional} causes a null value if the
function has failed, or otherwise, it returns the contained value,
that is, \texttt{M2Solution} in our case.
The function fails if the input is incorrect or the function has
eventually failed to find a minimum. Once the calculation of the
function is successful, the result can be extracted by the
\texttt{value} method of \texttt{std::optional}.
\begin{lstlisting}
const auto m2sol = yam2::m2CCSQP(input.value());
if (!m2sol) {
    std::cerr << "Failed.\n";
} else {
    std::cout << "M2CC = " << m2sol.value().m2() << '\n'
              << "solution:\n"
              << "  k1: " << m2sol.value().k1() << '\n'
              << "  k2: " << m2sol.value().k2() << '\n';
}
\end{lstlisting}
As can be seen in the code snippet, the \texttt{M2Solution} class
contains three methods: \texttt{m2} for the $M_2$ value, \texttt{k1}
and \texttt{k2} for the $M_2$ solution to the invisible particle
momenta.
All the functions and classes are in the namespace of \texttt{yam2}.
Once the minimization is successful,
the $M_2$ solution can be used to calculate other collider
variables such as $M_\text{AT}$ in Ref.~\cite{Park:2020rol}, after
converting it into a suitable object. For example,
the instance of \texttt{TLorentzVector} in \texttt{ROOT} can be
constructed from the solution as follows.
\begin{lstlisting}
const auto k1 = m2sol.value().k1(), k2 = m2sol.value().k2();
const TLorentzVector inv1{k1.px(), k1.py(), k1.pz(), k1.e()};
const TLorentzVector inv2{k2.px(), k2.py(), k2.pz(), k2.e()};
\end{lstlisting}

There are three inputs to the functions for calculating $M_2$.
The first one is an instance of \texttt{InputKinematics}, which is for
the particle momentum configuration of the given event. It can be
constructed by using the \texttt{mkInput} function,
\begin{lstlisting}
std::optional<InputKinematics> mkInput(
    const std::vector<FourMomentum> &as,
    const std::vector<FourMomentum> &bs,
    const TransverseMomentum &ptmiss, const Mass &minv);
\end{lstlisting}
Here \texttt{as} and \texttt{bs} correspond to the four-momenta of
the visible particles $a_i$ and $b_i$. See the decay topology given
in~(\ref{eq:pair_cascade}).
The convention is
\begin{equation}
  \texttt{as} = (\vb*{p}_{a_1}, \, \vb*{p}_{a_2}) , \quad
  \texttt{bs} = (\vb*{p}_{b_1}, \, \vb*{p}_{b_2}) .
\end{equation}
We stress that the order of the particle momenta should be set
with care since it is not checked by the program: $a_i$ must be
produced before having $b_i$ in the decay chain.
In addition to them, users have to insert the missing transverse
momentum and the invisible particle mass into \texttt{ptmiss} and
\texttt{minv}, respectively.
Schematic structures of the momentum classes are
\begin{align}
  \texttt{class}\,\,\, \texttt{FourMomentum}
  &= \texttt{FourMomentum}\,\, (E, \, P_x, \, P_y, \, P_z), \nonumber\\
  \texttt{class}\,\,\, \texttt{TransverseMomentum}
  &= \texttt{TransverseMomentum}\,\, (P_x, \, P_y).
\end{align}
See \texttt{momentum.h} for the class definitions of
\texttt{FourMomentum}, \texttt{TransverseMomentum}, and \texttt{Mass}.
The input momentum configuration should be validated before
substituting it into the functions for calculating $M_2$.
An example code snippet using the \texttt{mkInput} is given below.
\begin{lstlisting}
const auto input =
    yam2::mkInput({a1, a2}, {b1, b2}, ptmiss, yam2::Mass{m_invis});
if (!input) {
    std::cerr << "Invalid input.\n";
}
const auto m2sol = yam2::m2CCSQP(input.value());
\end{lstlisting}
The other optional inputs to the \texttt{m2CCSQP} function in the
above are the tolerance (\texttt{eps}) and the maximal number of iterations
(\texttt{neval}).
These will be set to the default values defined in \texttt{yam2.h} unless
users supply any input. In the current version of \texttt{YAM2}, their
default values are $\texttt{EPS} = 10^{-3}$ and $\texttt{NEVAL} = 5000$.
We recommend users to read the example analysis code enclosed with
\texttt{YAM2}, \texttt{examples/m2.cc}, before starting to write their
analysis code for the $M_2$ variables.

Finally, we show an example command for building an analysis code
using \texttt{YAM2}. Supposing that the name of the analysis code is
\texttt{m2.cc} and the path to \texttt{YAM2} is
\texttt{/usr/local}, the command is as follows.
\begin{center}
\begin{varwidth}{\linewidth}
\begin{verbatim}
c++ -o m2.exe m2.cc -I/usr/local/include/YAM2 \
    -L/usr/local/lib -lYAM2 -lnlopt
\end{verbatim}
\end{varwidth}
\end{center}

\section{Summary and outlook}

\noindent
The $M_2$ variables are an extension of $M_{T2}$ by promoting the
transverse masses to Lorentz-invariant ones and making explicit use of
on-shell mass relations.
Depending on the on-shell mass relations, the $M_2$ variables have a
hierarchical structure, which results in higher event density of the
distribution near the parent particle mass.
Finding the $M_2$ value and solution corresponds to solving a
constrained minimization problem.

Due to the lack of general analytic expression for the $M_2$ value and
solution, the calculation relies on numerical minimization
algorithms.
Up to the present time, the only publicly available software package for
calculating $M_2$ is \texttt{OPTIMASS}, where the augmented Lagrangian
method with Migrad and Simplex algorithms has been employed.
We note that there exist various numerical methods for solving
constrained optimization problems.
Among them, we have chosen the sequential quadratic programming method
and the derivative-dependent BFGS algorithm.
The method has been codified by using the implementation of the
numerical algorithms in the \texttt{NLopt} library.
The new library, \texttt{YAM2}, also includes other numerical
algorithms for calculating $M_2$, such as the augmented Lagrangian
method with the BFGS update.

By using \texttt{YAM2}, we performed a benchmark study for checking
the performance of the numerical algorithms.
It turned out that the sequential quadratic programming method
correctly identified the local minimum for $M_2$, and it is more
efficient than the other numerical methods, as well as
\texttt{OPTIMASS}.
We release \texttt{YAM2} as publicly available free software, to help
physicists interested in the $M_2$ variables for applying them to
physics analyses.

There are many rooms for upgrading and adding more features to
\texttt{YAM2}. We list a few of them:
\begin{itemize}
\item As supposed in Ref.~\cite{Cho:2014naa}, the $M_2$ variables can
  be defined for various subsystems of visible particles.
  It is also possible to calculate $M_2$ for different subsystems
  using \texttt{YAM2}, but the interface is not very transparent.
  We will improve the interface in the upcoming release of the
  upgrade.
\item Other than the sequential quadratic programming, one of the most
  popular algorithms for constrained optimization problems is the
  interior-point method. As there exists a publicly available software,
  \texttt{Ipopt}~\cite{cite:ipopt}, for the interior-point method, it
  would be straightforward to test the method.
\item Another interesting variable related with $M_2$ is
  $M_{2\text{Cons}}$, which is defined as~\cite{Konar:2015hea,
    Konar:2016wbh}:
  \begin{align}
    M_{2\text{Cons}} \equiv
  &~ \min_{\vb*{k}_{1}, \, \vb*{k}_{2} \in \mathbb{R}^3}
  \Big[ \max \Big\{ M \left( p_{1},
      \, k_{1};\, M_\chi \right), \,
    M \left( p_{2}, \, k_{2};\, M_\chi \right) \Big\} \Big]
    \nonumber\\
  &~ \text{subject to}\,\, \left\{\,
    \begin{aligned}
      \vb*{k}_{1T} + \vb*{k}_{2T}
      &= \slashed{\vb*{P}}_T , \\
      (p_1 + p_2 + k_1 + k_2)^2
      &= M_X^2 .
    \end{aligned} \right .
  \end{align}
  It can be useful for measuring the masses of on-shell intermediate
  particles produced in a pair from a resonance,
  \begin{equation}
    X \longrightarrow Y + \bar Y
    \longrightarrow v_1 (p_1) \chi (k_1) + v_2 (p_2) \bar \chi (k_2) ,
  \end{equation}
  where the resonance mass $M_X$ is known {\em a priori}. The $M_{2\text{Cons}}$
  distribution is bounded from above by $M_Y$.
  We can implement the $M_{2\text{Cons}}$ variable in the same way as
  in $M_2$.
\item Since the source code of \texttt{YAM2} is written in \CC, it can
  directly be used in analysis codes written in \CC\@.
  However, we expect that providing a C wrapper for \texttt{YAM2} will
  greatly help to use it in the codes written in the other
  programming languages through foreign function interface.
 \end{itemize}

\section*{Acknowledgments}

The author is grateful to Doyoun~Kim and Seodong~Shin for their useful
comments on the manuscript.
This work was supported by IBS under the project code, IBS-R018-D1.

\bibliography{yam2}

\end{document}